%% file: kTight.tex
\documentclass[11pt]{article}
\usepackage{fullpage}
\usepackage{amsmath}
\usepackage{algorithm,algorithmic,ifthen,url,graphicx} 
\usepackage{times}

\newenvironment{proof}{\par{\bf Proof:}}{\hfill$\square$ \par}

\newcommand{\qedF}{}

\usepackage{color}

\input macrosFlo.tex

\newcommand{\demandSet}[1]{\mathcal{D}^{#1}}

\title{Sequential item pricing for unlimited supply} 
\author{Maria-Florina Balcan\thanks{College of Computing, Georgia Institute of Technology. \texttt{$\{$ninamf,florin$\}$@cc.gatech.edu}. 
This work was supported in part by NSF grant CCF-0953192, by ONR grant N00014-09-1-0751, and 
by AFOSR grant FA9550-09-1-0538. 
} \and Florin Constantin$^{\ast{}}$} 
\date{}

\newcount\Comments  
\definecolor{darkgreen}{rgb}{0,0.7,0}
\definecolor{purple}{rgb}{1,0,1}
\newcommand{\kibitz}[2]{\ifnum\Comments=1\textcolor{#1}{#2}\fi}

\newcommand{\QA}[1]{\textbf{\emph{#1}}}

\newcommand{\revApproxFactorMath}[3]{\ensuremath{C_{#1,#2,#3}}}

\begin{document} 
\maketitle

\newcommand{\RndMidsAlgo}{\ensuremath{\mathbb{A}}} 
\renewcommand{\vec}[1]{\ensuremath{\mathbf{#1}}}
\newcommand{\mbF}{\mathbf{F}}
\newcommand{\myopic}{forward-myopic}

\newcommand{\buyers}{\ensuremath{1..m}}

\renewcommand{\labelitemi}{$\bullet$}

\newcommand{\abc}{\ensuremath{\{a,b,c\}}}



\newcommand{\nOrmn}{mn}


\vspace{-2\baselineskip}
\begin{abstract}
 {
$\!\!$
We investigate the extent to which 
price updates can 
increase the revenue 
of a seller with little prior information on demand.} 
We study prior-free revenue maximization for a seller with unlimited supply of $n$ item types facing $m$ myopic buyers  present for $k\!<\!\log n$ days. 
%
%
For the static ($k=1$) case, Balcan et al.~\cite{BalcanBM08ItemPfRM} show that one random  item price (the same on each item) yields revenue 
within a $\Theta(\log m + \log n)$ factor of optimum and this factor is tight. 

We introduce \emph{hereditary maximizers}, a novel property regarding buyer valuations 
that is sufficient for a significant improvement of the approximation factor in the dynamic ($k>1$) setting. 
The hereditary maximizers property limits complementarities among items  and is satisfied by any multi-unit or gross substitutes valuation. 
Our main result is a non-increasing, randomized, schedule of $k$ item prices, the same on each item, with expected revenue 
within a $O(\frac{\log m + \log n}{k})$ factor of optimum for private valuations with hereditary maximizers. 
%
This factor
is \emph{tight} (modulo a constant factor): 
we show that any pricing scheme over $k$ days has a revenue approximation factor of at least $\!\frac{\log m+\log n}{3k}$. 
We obtain analogous matching lower and upper bounds of $\Theta(\frac{\log n}{k})$ if all valuations have the same maximum. 
 We expect the technique we develop to prove our upper bound  to be of broader interest; for example, it can be used to significantly improve over the result of~\cite{AkhlaghpourGHMMN09OptimalIPwPNE}. 
 
We also initiate the study of revenue maximization given allocative  \emph{externalities} (i.e. influences) between buyers with combinatorial valuations. 
We provide a rather general model of positive influence of others' ownership of items on a buyer's valuation.  
For affine, submodular externalities and valuations with hereditary maximizers 
we present an influence-and-exploit~\cite{HartlineMS08OptimalMSoSN} marketing strategy based on our algorithm for private valuations.
 This strategy preserves our 
 approximation factor, despite an affine increase (due to externalities) in the optimum revenue. 
\end{abstract}

\vspace{-1.5\baselineskip}
\section{Introduction} 
\vspace{-0.5 \baselineskip}

In most transactions, prices are set on items (and not on bundles of items) to  simplify buyers' and sellers' decisions. 
Arguably, a seller's prevalent objective is to maximize revenue $\!$--$\!$ this basic problem has received tremendous attention in the optimization  literature. 
The seller often has poor information about  uncertain demand, 
\mbox{but still   desires a mechanism  robust to errors or omissions in its data.}

Given a set of buyers and their valuations for bundles, the optimum revenue is the total value of the optimal allocation. 
Posted prices that obtain such a high revenue do not usually exist, because buyers' valuations are private and may be quite complex. 
A standard compromise~\cite{BalcanBM08ItemPfRM,BansalCCRSS10DynamicPfIB,ChakrabortyHK09NonUniformPSfRM} 
 is then to aim for revenue that is at least (possibly in expectation) a fraction $1/c$ of the optimum for \emph{any} set of buyers, or, more formally, to design algorithms with a low revenue approximation factor $c>1$. 

We focus on \emph{unlimited supply}~\cite{BalcanBM08ItemPfRM,BansalCCRSS10DynamicPfIB},  
a setting relevant to digital media (e.g. DVDs or software programs).  
A seller has unlimited supply
of $n$ item types 
if the marginal cost of producing an additional copy of any item is negligible.  
For unlimited supply, 
the highest possible revenue equals the sum of the maxima of buyers' valuations. 
Assuming that the seller only knows an upper bound on the $m$ buyers' 
arbitrary valuations, Balcan et al.~\cite{BalcanBM08ItemPfRM} provide 
a \emph{one-shot} randomized  price (the same for each item) that yields revenue at least a $\Theta(1/({\log m + \log n}))$ fraction of the optimum; they also show that this factor is tight for $m\!=\!1$ buyer.

In practice buyers purchase more than once from the same seller. 
It is then natural to investigate improved approximation factors if all buyers are present for $k\!<\!n$ time periods, that we call \emph{days}. 
%


Like~\cite{BalcanBM08ItemPfRM}, 
we price all items equally, i.e. we use \emph{linear uniform} prices. 
This involves the least price discrimination possible under static pricing: no buyer or item is favored over another. 
In fact, some online movie retailers (e.g. iTunes) have very limited variability in prices -- iTunes offers only two prices for movies, older movies having a discount. 
Prices must typically be decided before observing demand. 
In a dynamic setting like ours,  the seller may update prices ``on the fly'' based on realized demand. For simplicity we only consider, like~\cite{BalcanBM08ItemPfRM,BansalCCRSS10DynamicPfIB}, price sequences decided ahead of time, but only revealed gradually to buyers. 

A buyer starts out with no items and accumulates them over time. 
We assume that any buyer is \emph{\myopic}, 
i.e. purchases a preferred set in each day without reasoning about future price reductions. If buyers 
could wait for the lowest price $p$ in a sequence $\calP$, $\calP$  would be just as effective as $p$.

We are now ready to state the central question in this paper
\vspace{-0.25\baselineskip}
\begin{center}
\QA{Question}: \emph{What revenue approximation factor \revApproxFactorMath{m}{n}{k}\ 
 is achievable with \\
$m$ \myopic\ buyers,
$n$ items in unlimited supply and $k$ equal item prices?} 
\end{center}

Balcan et al.~\cite{BalcanBM08ItemPfRM}'s results 
can be re-stated as $\revApproxFactorMath{m}{n}{1} \!=\! O(\log m \!+\! \log n)$ 
and  $\revApproxFactorMath{1}{n}{1} \!=\! \Theta(\log n)\!$. 

We provide a general lower bound and an upper bound on the revenue factor achievable (we introduce in detail these results and define these valuations shortly). 


\begin{center}  \ \ \QA{Answer}: 
    $\begin{cases}
\text{ \it Even for public concave multi-unit valuations,  }
\!\frac{\log m+\log n}{3k} \!\leq\! & \!\!\!\!\revApproxFactorMath{m}{n}{k}\!
\\
\text{\emph{For private valuations with hereditary maximizers,}} 
& $\!\!\!\!$\revApproxFactorMath{m}{n}{k} \!=\! O(\frac{\log m + \log n}{k})\!
\end{cases}$
\end{center}
\vspace{-0.25\baselineskip}


We show that no scheme with $k$ successive prices can approximate revenue to a factor less than $\frac{\log m+\log n}{3k}$, even for 
 concave multi-unit (i.e. that do not differentiate items)  valuations. Such valuations are among the most basic combinatorial valuations, rendering our lower bound quite powerful. 

Our main result however is a positive one, providing a matching upper bound. 
We show that generating $k$ independent random prices and offering them in decreasing order 
(with the same price on each item in a given day)
 approximates revenue to no worse than a $\Theta(\frac{\log m + \log n}{k})$ fraction, thus improving~\cite{BalcanBM08ItemPfRM}'s approximation by a $\Theta(k)$ factor. 
Our technical contribution is to generalize a guarantee on the expected  profit from one random price to $k$ such prices. While the bound for one price uses a standard technique for worst-case bounds, the only improvement for general $k$ that we are aware of (by Akhlaghpour et al.~\cite{AkhlaghpourGHMMN09OptimalIPwPNE}) 
is exponentially worse than ours as their recursive construction only yields a $\Theta(\log k)$ factor improvement. 
We connect revenue from a valuation $v$ with the joint area of $k$ rectangles, determined by prices, under $v$'s demand curve $F$. While each such rectangle covers in expectation a logarithmic fraction of the area under $F$, we are able to limit the overlaps of rectangles by carefully analyzing the $k$ prices as order statistics. 
If all valuations have the same maximum (or obviously if $m = n^{O(1)}$), then we can improve our two bounds to $\Theta(\frac{\log n}{k})$.

Our upper bound (in particular our connection between revenue and area under demand curve covered by price-based rectangles) relies on the natural sufficient condition (that we identify) of buyer valuations having \emph{hereditary maximizers} (HM). 
The HM property essentially states that an algorithm greedily selecting items by their marginal value 
has at each step a set of maximum value. 
In particular, multi-unit (not necessarily concave) valuations and 
gross substitutes valuations (a classical model in economics, see e.g.~\cite{BikhchandaniO06AscendingPVA,GulS99WalrasianEwGS}) 
have HM. 
We discuss these and other natural HM valuations in Section~\ref{sec:HMnGSEx}. 

Submodular valuations may not have  HM, leading to a counter-intuitive phenomenon: the revenue from offering a high price followed by a low one may be less than the revenue from the low one only. 
Consider three movies: 
a very good science-fiction (S) one and an animation (A) movie and a drama (D), both of slightly inferior quality. 
A typical family prefers S to A or D, 
but A and D (for variety) to any other pair; the family does not strategize about price schedules. 
If a greedy movie retailer starts with high prices and  reduces them afterwards (on all movies) then, despite good revenue on S, it loses the opportunity of more revenue by selling A and D instead. 
This (submodular) valuation, formalized in Section~\ref{sec:seqPr}, does not have HM. 
For HM valuations however, this counter-intuitive reversal does not occur,  which is critical, as we show, for good sequential revenue. 


In the final section of the paper, we allow a buyer's valuation   
 for a set of items to depend on the others' ownership of copies of (possibly different) items. 
 %
In economic language, others exert an allocative\footnote{As opposed to 
\emph{informational} or \emph{financial} externalities~\cite{JehielM06AllocativeaIEiAaRM}, where a valuation depends on others' information (e.g. signals of the item's quality) or on their payments.} externality on a buyer's value. 
Movie distribution  services (e.g. Netflix) exploit such effects, allowing users to befriend each other and to observe which movies they watched (and their rating). 

We introduce a prior-free model of externalities and extend our algorithm for private valuations. 
Our model departs from existing~\cite{AkhlaghpourGHMMN09OptimalIPwPNE,HartlineMS08OptimalMSoSN} revenue maximization problems in the presence of externalities in two aspects, 
that, in our view, allow for more generality. 
First, there is more than one item type for sale, which requires a new language for expressing externalities. We introduce such a language that extends~\cite{AkhlaghpourGHMMN09OptimalIPwPNE,HartlineMS08OptimalMSoSN}; it allows a buyer to express a positive, affine influence by others that is monotone and submodular in their bundles. 
Second, we assume only certain properties of the valuation functions, as opposed to values drawn from a (known) distribution. 
We obtain an algorithm with the same approximation ratio as without externalities, despite an affine increase in the optimum. 
Our algorithm is an influence-and-exploit (IE) strategy, introduced by Hartline et al.~\cite{HartlineMS08OptimalMSoSN}. 
In an IE strategy, a set $S$ of buyers is given some items for free and then other customers are charged a price that exploits other owners' (a superset of  $S$) influence on the items' value. 


\newcommand{\rwh}[1]{} 
\textbf{Related work}. 
\rwh{Unlimited supply}
Revenue maximization has been studied with and without priors on buyer valuations and with limited or unlimited supply. The prior-free, unlimited supply domains studied have been less  general than the one in this paper. 
Balcan et al.~\cite{BalcanBM08ItemPfRM} present structural results for 
one-shot unlimited supply pricing and arbitrary valuations no higher than $H$. They achieve a tight $\Theta(\log m \!+\! \log n)$-factor revenue approximation via  a single random price. 
%
Bansal et al.~\cite{BansalCCRSS10DynamicPfIB} study
sequential pricing for unlimited supply of one item type and buyers  with  values in $[1,H]$ and arrival--departure intervals. They obtain almost matching upper and lower  bounds on the approximation factor: $O(\log H)$ for deterministic schemes and $O(\log \log H)$ for randomized schemes. 
Guruswami et al.~\cite{GuruswamiHKKKM05OnProfitMEFP} present static logarithmic revenue approximations via envy-free pricing for two settings:  single-minded buyers for unlimited supply
and 
unit-demand buyers for limited supply. 
The only dynamic version they consider concerns gradually-expiring items and buyers with arrival-departure intervals. 
Chakraborty et al.~\cite{ChakrabortyHK09NonUniformPSfRM} 
achieve a $O(\log^{2}n)$-factor revenue approximation   via dynamic  equal item prices, improving the limited supply factor in~\cite{BalcanBM08ItemPfRM}.  
Unlike us, they consider a limited supply setting and
assume that buyers are impatient and have  subadditive valuations. 
%
%
Hajiaghayi et al.~\cite{HajiaghayiKP04AdaptiveLSOA} obtain constant factor revenue approximations via adaptive pricing 
for limited supply of a good and 
buyer values from a distribution. 
Chawla et al.~\cite{ChawlaHMS10MultiparameterMDaSPP} present constant-factor revenue approximation schemes via sequential posted prices  in prior-based domains. 

\rwh{Externalities in social networks} 
While externalities are natural and well-studied in social networks~\cite{Kleinberg07CascadingBiNAaEI}, 
the corresponding revenue maximization problem has been recently introduced by Hartline et al.~\cite{HartlineMS08OptimalMSoSN}, who  investigate approximation via single-item distribution-based influence-and-exploit marketing strategies. Akhlaghpour et al.~\cite{AkhlaghpourGHMMN09OptimalIPwPNE} 
study this problem for a seller that cannot use price discrimination amongst buyers. 
\rwh{Externalities in Econ} 
Jehiel and Moldovanu~\cite{JehielM06AllocativeaIEiAaRM} find that many classical results no longer hold 
when externalities (allocative or informational) arise in auctions.   

\textbf{Paper structure}. %
%
After introducing notation 
in Section~\ref{sec:noExty},  
we review known bounds on $\revApproxFactorMath{m}{n}{k}$ and provide a new lower bound in Section~\ref{sec:existLower}. 
Section~\ref{sec:HM} analyzes hereditary maximizers, a property  of valuations leading to a  $\revApproxFactorMath{m}{n}{k}$ upper bound established in Section~\ref{sec:HMApprox}. 
Finally, in Section~\ref{sec:extys}, we model externalities, where a buyer's valuation depends on  others' items, 
\mbox{and extend Section~\ref{sec:HMApprox}'s approximation.}

\vspace{-0.5\baselineskip}
\section{Preliminaries}
\label{sec:noExty}
\vspace{-0.5\baselineskip}


We consider a seller with $n$ item types in unlimited supply. The seller can thus profit from selling copies of an item at any price 
but aims to maximize its revenue. 
The seller has $k<n$ sale opportunities called \emph{days}. There are $m$ customers with quasilinear utilities present in all $k$ days. 
Customers have valuations over bundles of items (not more than one per  type); we denote a generic such valuation\footnote{\mbox{$\!\!$A valuation is constant over time, apart from Section~\ref{sec:extys}, where it  varies with others' items.}} by  
$v_{i} \!:\! 2^{1..n} \!\!\to\! \reals.\!$ and its maximum by $H_{i}$. 
We assume that the seller knows only the highest maximum across  customers $H \!= \max_{i} H_{i} \!= \!\max_{i \in \buyers, S \subseteq 1..n} \! v_{i}(S)$. 

 We treat static pricing first and then dynamic pricing in Section~\ref{sec:seqPr}. 
We only use the simplest form of pricing, with no item or buyer discrimination. 
A price vector $\vec{p}\!\in\!\reals^{n}$ is \emph{linear uniform}\footnote{Different (non-uniform) item prices are also (e.g.~\cite{BikhchandaniO06AscendingPVA}) called linear prices.} if 
$p_{j} \!=\! p, \all j \!=\!1..n$.  

\noindent
Given a price vector, a customer buys a preferred (utility-maximizing) bundle. 
\begin{Definition}
For price vector $\vec{p}\in\reals^{n}$, the \emph{demand correspondence}~\cite{GulS99WalrasianEwGS} $\demandSet{}_{v}(\vec{p})$ 
of valuation $v$ is the set of utility-maximizing bundles at prices $\vec{p}$: 
\vspace{-0.5\baselineskip}
\begin{align}
\label{eq:demandSet} \tst
 \demandSet{}_{v}(\vec{p}) = \argmax_{S \subseteq 1..n} \{ v(S) - \sum_{j \in S} p_{j}\}
\end{align}
For linear uniform price $\vec{p} \!=\! p \cdot \vec{1}$, let  $\demandSet{}_{v}(p) \!=\! \demandSet{}_{v}(\vec{p})\!$ and $F_{v}(p) \!=\! \min_{S \in \demandSet{}(p\cdot \vec{1})} \!|S|\!$ be the least number of items in a bundle demanded (by valuation\footnote{Except for Section~\ref{sec:extys}, $v$ will be clear from context and omitted from $\demandSet{}$ and $F$.} $v$) at prices $p$. 
\label{def:demandSetF}
\end{Definition}
\vspace{-\baselineskip}

As one would expect, a higher price cannot increase the least  quantity bought. 
\vspace{-0.5\baselineskip}
\begin{Lemma} (Balcan et al.$\!$~\cite{BalcanBM08ItemPfRM}) 
For an arbitrary  valuation $v$ and $p \!>\! p'\!$, 
$F(p) \!\leq\! F(p')\!$. 
\label{stmt:Fdecrp}
\end{Lemma}
\vspace{-0.5\baselineskip}

\newcommand{\areaF}[1]{\mathrm{A}_{#1}}

\vspace{-\baselineskip}
\subsection{Sequential pricing}
\label{sec:seqPr}
\vspace{-0.5\baselineskip}

Assume the seller offers equal item  prices 
$r^{d} \!\in\! \reals_{+}$  in day  
$d \!=\! 1..k$, with $r^{1}\! \!> \!\ldots\! >\! r^{k}\!$. 
We now define customer behavior over time, starting with no items before day $1$.

$\!$We model any buyer as \emph{\myopic}:   
assume that before day $d$ he buys sets 
$S_{1}, \dots, \!S_{d-1}$. His utility for items 
$\!S \!\subseteq\! 1..n \!\setminus\! (S_{1} \!\cup\! \ldots\! \cup S_{d-1}\!)\!$ 
he does not own is 
\begin{align}
\vspace{-0.5\baselineskip}
\!\!\!\!u_{d, \dots, 1}(S_{1}, \!\dots\!, S_{d-1},S,r^{1}\dots r^{d}) = 
v(S_{1} \!\cup \!\dots\! \cup\! S_{d-1} \!\cup\! S) \!-\! (\tst\sum_{l=1}^{d-1} r^{l} |S_{l}|) \!-\! r^{d} |S| 
 \label{eq:seqUt}
\vspace{-0.5\baselineskip}
\end{align}
i.e. a customer does not  anticipate price drops but does take into account past purchases (accumulating items) and payments to decide a utility-maximizing set $S$ to buy today. 
In this model, a customer buys nothing in a day  where the price increases\footnote{ {We sketch a proof for $d\!=\!2$: let $r^{1} < r^{2}$ and  $S_{i}$ bought in day  $i\!=\!1,2$ with $S_{1} \!\cap\! S_{2} \!=\! \emptyset$.  Suppose $S_{2} \!\neq\! \emptyset$; then $v(S_{1} \!\cup\! \emptyset) \!-\! r^{1}|S_{1}| \!-\! r^{2}|\emptyset| 
\!\leq\! v(S_{1} \cup S_{2}) \!-\! r^{1}|S_{1}| \!-\! r^{2}|S_{2}| \!<\! v(S_{1} \cup S_{2}) \!-\! r^{1}|S_{1}| \!-\! r^{2}|S_{2}|$, i.e. $S_{1} \cup S_{2}$ is preferred to $S_{1}$ at price $r^{1}$, contradiction.}}, hence our focus on decreasing price sequences:  
the seller starts with a high price and then gradually reveals  discounts, a common retail practice.  

$\!$Following Def.~\ref{def:demandSetF}, we 
denote 
 preferred bundles outside $S_{1} \!\cup \!\ldots \cup S_{d-1}$ at 
$r^{d}$ 
by 
\vspace{-0.5\baselineskip}
 \begin{align*}
 \demandSet{S_{1}, \dots, S_{d-1}}_{v}(r^{1} \dots r^{d}) = 
\!\argmax_{S \subseteq 1..n \!\setminus\! (S_{1} \cup \dots \cup S_{d-1})}\!u_{d, \dots, 1}(S_{1}, \dots, S_{d-1}, S, r^{1} \dots r^{d})
\end{align*} 
 \vspace{-1.5\baselineskip}

We briefly consider incentive properties before focusing on revenue only. 

 {\textbf{Incentive considerations.} A buyer's utility cannot decrease in any day: there is always the option of not buying anything. Thus, any sequence of prices defines an individually rational mechanism. 
Furthermore, within a day, as each buyer faces  the same prices, buyers have no envy and no profitable item swaps. }
 

\newcommand{\revenue}[1]{\mathrm{Rev}_{#1}}

Our goal is revenue maximization via (possibly randomized) price sequences decided ahead of time (but only revealed gradually to buyers). 
\begin{Definition}
A pricing scheme $\calP$ is a sequence\footnote{
We can expand, without changing revenue,
any shorter sequence $\calP$ to $k$ prices by appending  to $\calP$ copies of its last price.} of
$k$ (possibly random) decreasing prices. 
$ \revenue{\calP}(v_{1}, \!\ldots\!, v_{m})$ denotes $\calP$'s  revenue (in expectation for randomized $\calP\!$), for valuations $v_{1},\!\ldots\!,v_{m}\!$ and least favorable tie-breaking decisions by buyers. 
\end{Definition}
\vspace{-0.5\baselineskip}

A standard~\cite{BalcanBM08ItemPfRM,HartlineMS08OptimalMSoSN} revenue benchmark is customers' total willingness to pay. 
We study worst-case guarantees, that hold regardless of buyer valuations.

\vspace{-0.5\baselineskip}

\begin{Definition}
A (possibly randomized) pricing scheme $\calP$ is a \emph{$c$-revenue approximation} (where $c\geq1$) if  
$\sum_{i\in \buyers} \max_{S \subseteq 1..n} v_{i}(S)$
$\leq$\footnote{This benchmark is at least $\!\max_{S \subseteq 1..n} \!\sum_{i\in \buyers} $\!$v_{i}(S)$, $\!$i.e.$\!$ the highest joint value of a set.}  
$ c \cdot \expec[\revenue{\calP}(v_{1}, \dots, v_{m})] $
for all valuations $v_{1} \dots v_{m}$,  
where the expectation is taken over $\calP$'s random choices. 
\end{Definition}

Recall that  this paper's central question is to assess what revenue approximation factors $\revApproxFactorMath{m}{n}{k}$ are achievable. 
Clearly, $\revApproxFactorMath{m}{n}{k+1} \!\leq\! \revApproxFactorMath{m}{n}{k}$ and $\revApproxFactorMath{m}{n}{k} \!\leq\! \revApproxFactorMath{m+1}{n}{k}$.  
The next section  formally  states known values of $\revApproxFactorMath{m}{n}{k}$ for particular $(m,n,k)$ triples, provides some intuition for sequential pricing and presents our lower bound $\revApproxFactorMath{m}{n}{k} = \Omega(\frac{\log m+\log n}{k})$.  The upper bound $\revApproxFactorMath{m}{n}{k} = O(\frac{\log m+\log n}{k})$ 
is presented 
in Section~\ref{sec:HMApprox}.

\section{Existing bounds for $\revApproxFactorMath{m}{n}{k}$ and a new lower bound}
\label{sec:existLower}
\vspace{-0.5\baselineskip}

We start by reviewing known or basic results on $\revApproxFactorMath{m}{n}{k}$  for  simple instances of our setting: first for one-shot 
pricing, then for one item 
only and finally for public valuations, which will also yield a new lower bound. 

Motivated by~\cite{BalcanBM08ItemPfRM}'s bounds (re-stated in Lemma~\ref{stmt:logBBM08} below) and 
other worst-case results~\cite{BansalCCRSS10DynamicPfIB}, 
we use 
prices 
\vspace{-0.5\baselineskip}
\[
q_{l} = H/2^{l} \text{ for } l \geq 0 
\]
%
%

\vspace{-0.5\baselineskip}
Algorithm \textsc{Random}$_{D}^{H}$ 
 outputs one-shot price $q_{l}$ where the scaling exponent $l$ is chosen uniformly at random in $0..D\!-\!1.\!$ Despite its simplicity,  \textsc{Random}$_{D}^{H}$ is quite effective in general and as effective as any other algorithm for one buyer. 

%
 \begin{Lemma}\cite{BalcanBM08ItemPfRM}
 \label{stmt:logBBM08}
For\footnote{This paper only uses  base $2$  logarithms.} $t \!=\! 1\!+\! \log m \!+\! \log n$,  
\textsc{Random}$_{t}^{H}\!$ 
is a $4t$-revenue approximation. 
For one buyer, i.e. $m\!=\!1$, this factor is tight (modulo a constant factor). 
Thus, $\revApproxFactorMath{m}{n}{1} \!=\! O(\log m \!+\! \log n)$ 
and  $\revApproxFactorMath{1}{n}{1} \!=\! \Theta(\log n)\!$. 
\end{Lemma}
\vspace{-0.5\baselineskip}
%

%


Bansal et al.~\cite{BansalCCRSS10DynamicPfIB} study sequential revenue maximization for unlimited supply of one item (as in our setting, buyers are interested in  one copy only). 
While they assume, as we do, that values are at most $H$, 
they also assume, unlike us, that values are at least $1$. 
They essentially show that $k\!=\!1\!+\!\log H$ days are enough for obtaining at least half of the maximum revenue. 
More precisely, 
\textsc{Random}$_{1+\log H}^{H}$ is a $2$-approximation to revenue, i.e. 
$\revApproxFactorMath{m}{1}{1+\log H} \!= \!2$ for such values. 


%
For the rest of this section we assume that the seller fully  knows buyers' valuations. 
This strong assumption will allow us to understand two special settings. 


The first setting is concerned with one buyer $m\!=\!1$  with a known monotone ($H\!=\!v(1..n)$) valuation $v$ and many days $k\!=\!n$. In this setting, full revenue can be obtained from  $v$ i.e. 
$\revApproxFactorMath{1}{n}{n}=1$.  
 Let set $S_{d}$ be bought in day $d=0..n$, with $S_{0} = \emptyset$. Let item $\sta{i}_{d} \in \argmax_{i \not\in S_{d}} v(S_{d}\wi{i})$ with the highest marginal value given $S_{d}$. 
 Then, at price $r^{d} \!=\! v(S_{d}\wi{\sta{i}_{d}}) \!-\! v(S_{d})$ in day $d \!\in\! 1..n$, exactly\footnote{Here we also assume $\!$$v$ submodular; if not, we can set $r^{d}\!$ to $v$'s steepest slope at $S_{d}$.} 
 item $\sta{i_{d}}$ is bought (ignoring ties). The sum of all days' revenues telescopes to $v(1..n)\!=\!H$. 

The second setting yields the first half of our answer to this paper's central question: 
a lower bound on achievable 
$\revApproxFactorMath{m}{n}{k}$. This lower bound shows 
the effect of limited ($k$) price updates and buyer differences, 
even if  valuations are known and items are identical.  


\newcommand{\lognExpScaledStmt}
{Define $
v_{s}(x) = 
\begin{cases}
 x/2^{s-1}, &\text{$\!$if } x \leq 2^{s-1} \\
1, &\text{$\!$if } x > 2^{s-1}
\end{cases} 
\all 1\! \leq\! s \leq \! N\!+\!1$ (for $N \!=\! \log n \in \integers$) 
to be $1+\log n$ concave multi-unit valuations, each with maximum $1$:  
 $v_{1}(x) = 1, \all x =1..n$ and $v_{N+1}(x) = x/2^{N} = x/n, \all x =1..n$. Then the revenue of any sequence of $k \!<\! \log n$ prices is at most ${2k}$. 
Thus, even if valuations have the same maximum, 
any $k$-day pricing algorithm must have a higher revenue  approximation factor than $\frac{1+\log n}{2k}$, even for $1+\log n$ buyers:  
$\frac{1+\log n}{2k} \leq \revApproxFactorMath{1+\log n}{n}{k}^{\textsf{equal maxima}}$. 
In general, $\revApproxFactorMath{m}{n}{k} = \Omega(\frac{\log m+\log n}{k})$. 
}


%
\vspace{-0.5\baselineskip}
\begin{Theorem} [lower bound]
\lognExpScaledStmt
\label{stmt:lognExpScaled}
\end{Theorem}
\vspace{-0.5\baselineskip}
 Informally, each $v_{s}$ has a constant non-zero marginal value (MV) for one item in $[1/n, 1]$. A low price is effective for low MV buyers but could profit more from high MV buyers. A high price fails to sell any item to 
low MV buyers despite getting good revenue from high MV buyers. 
This reasoning extends to short ($k \!<\! \log n$) sequences of prices. 

\begin{proof}
The maximum revenue from $v_{1}..v_{N+1}$ is $N\!+\!1$. 
We claim that no sequence of $k$ prices can yield revenue above $2k$. 
Any such sequence 
can be replaced by  
 $1/2^{l_{1}}, 1/2^{l_{2}}, \dots, 1/2^{l_{k}}$ with $0 \!\leq\! l_{1} \!<\! l_{2} \!<\! \ldots\! l_{k} \!\leq\! N$ without decreasing revenue.\footnote{ {Any price (or price sequence) in $(1/2^{l+1}\!, 1/2^{l})$ for $l \!\in\! 1..N\!+\!1$ can be replaced by $1/2^{l}$ while preserving the quantity bought by any $v_{s}$ as $v_{s}$'s marginal values equal $1/2^{s-1}\!$.}}

A price $1/2^{l}$ presented to a valuation $v_{s}$ yields 
revenue $2^{s-1}/2^{l}$ otherwise (all $2^{s-1}$ items being bought)  if 
$s-1 \leq l$ (i.e. if $1/2^{l} \leq 1/2^{s-1}$)
and $0$ otherwise. 
Similarly, the revenue from a price pair $1/2^{l_{1}}, 1/2^{l_{2}}$ with $l_{1}<l_{2}$ is 
$0$ if $s-1 > l_{2}$,  
$2^{s-1}/2^{l_{1}}$ if $s-1 \leq l_{1}$ 
and $2^{s-1}/2^{l_{2}}$ if $s-1 \in (l_{1}, l_{2}]$. 
The total revenue is then
\vspace{-0.5\baselineskip}
\begin{align*}
\ & \tst \sum_{s=1}^{l_{1}+1} 2^{s-1}/2^{l_{1}} 
+ \sum_{s=l_{1}+2}^{l_{2}+1} 2^{s-1}/2^{l_{2}} 
+ \ldots
+ \sum_{s=l_{k-1}+2}^{l_{k}+1} 2^{s-1}/2^{l_{k}} \\
&= (2^{l_{1}+1} \!-\! 1)/2^{l_{1}} 
 + (2^{l_{2}+1} \!-\! 2^{l_{1}+1})/2^{l_{2}} 
+ \ldots
+ (2^{l_{k}+1} \!-\! 2^{l_{k-1}+1})/2^{l_{k}}  
\leq 2k \ \ \  \qedF 
\end{align*}

Let $m' = m\!-\!1\!-\!\log n$ and assume that $m' > 0$. 
Let unit-demand single-minded valuations $w_{t}$ for $t = 1..m'$ such that $w_{t}(x) = w_{t}(1) = \frac{1}{t}, \all x = 1..n$. 
Then any $k$ prices can obtain total revenue at most $k$ from valuations $w_{1} .. w_{m'}$ whose optimal revenue is $\Theta(\log m')$.

For the $m \!=\! m'\!+\!1\!+\!\log n$ valuations $v_{s}$ and $w_{t}$ together, 
we get that no set of $k$ prices obtains more than $3k$ revenue, 
but optimum is $\Theta(\log n) + \Theta(\log(m \!-\! \log n)) \!=\! \Theta(\log m \!+\! \log n)$ i.e., $\revApproxFactorMath{m}{n}{k} \!=\! \Omega(\frac{\log m+\log n}{k})$. 
 \end{proof}

For $k\!=\!1$, we get $\revApproxFactorMath{m}{n}{1} \!=\! \Theta(\log m \!+\! \log n)$, showing that Balcan et al.~\cite{BalcanBM08ItemPfRM}'s bound 
 $\revApproxFactorMath{m}{n}{1} \!=\! O(\log m \!+\! \log n)$ is tight (they only prove it for $m=1$).


After  this section we will provide positive results only. 
In preparation however, we need another piece of bad news  highlighting the importance of sequential consistency. 

Even the seemingly innocuous assumption of decreasing prices can hurt revenue. 
We now provide a submodular valuation consistent with the movie example in the introduction 
and find that the revenue from a high, followed by a low, price may be below that of the better single price. 
\begin{Example}
Let $a$ be the science-fiction movie, and $b,c$ be the animation and drama. 
 Define a valuation $v$ by 
$v(a)\!=\!3, v(b)\!=\!v(c)\!=\!2.1, v(a,b)\!=\!v(a,c)\!=\!3.8, v(b,c)\!=\!v(a,b,c)\!=\!4.2$. 
For $r^{1} \!=\! 1.5$, 
$\demandSet{}(r^{1}) \!=\! \{\{a\}\}$ and for $r^{2} \!=\! 1, \demandSet{}(r^{2}) = \{\{b,c\}\}$. 
Neither $b$ or $c$ is worth \$$1$ given $a$: 
$\demandSet{\{a\}}(r^{1},r^{2})= \{\emptyset\}$. 
Less revenue (\${}$1.5$) is obtained from offering $r^{1}$ followed by $r^{2}$ than from $r^{2}$ alone (\${}$2$). 
\label{ex:noGSnoS1subsetS2}
\end{Example}
\vspace{-0.5\baselineskip}
%

In Example~\ref{ex:noGSnoS1subsetS2}, $v(1)>\max(v(2),v(3))$ and $v(23)>\max(v(12),v(13))$. 
The following example shows that 
revenue does not decrease for any two prices (and thus any price sequence)
for another monotone submodular non-GS valuation. Eq.~\eqref{eq:GSabcS} is violated, but unlike in Example~\ref{ex:noGSnoS1subsetS2}, only one of $v(1)$ and $v(23)$ is higher than the value of both other bundles of same size. 
 

Section~\ref{sec:HM} introduces
a valuation class for which the revenue from any  sequence of prices $r^{1} > \dots > r^{k}$ is at least that from the best $r^{i}$, in contrast to Example~\ref{ex:noGSnoS1subsetS2}. 
This property will yield revenue guarantees for our pricing scheme in Section~\ref{sec:HMApprox}.    

\vspace{-0.5\baselineskip}
\section{Hereditary maximizers}
\label{sec:HM}
\vspace{-0.5\baselineskip}

In this section we define hereditary maximizers, a new property of valuations 
and we show that it holds for�
a few classical domains of valuations including gross substitutes and multi-unit demand valuations. 
 In Section~\ref{sec:HMApprox} we will show that this property is sufficient for good sequential revenue. 


\vspace{-0.5\baselineskip}
\begin{Definition}
Valuation $v$ has \emph{hereditary maximizers} (HM) if 
given  any size $j$ 
value-maximizing bundle $S_{j}$, 
 one item can be added to it to obtain a size $j\!+\!1$ such 
 bundle.  
Letting 
$\calM_{j}^{v} \!\!=\! \tst\argmax_{|S|=j} v(S) $, $v$ has HM if
\begin{align}
\vspace{-0.5\baselineskip}
&\all n \!\geq\! j \!\geq\! 1, \all S_{j} \in \calM_{j}^{v}, \exists S_{j+1} \in \calM_{j+1}^{v} \text{ with } S_{j} \subset S_{j+1} \tag{HM} \\
\text{implying } & 
\all n \!\geq\! j' \!>\! j \!\geq\! 1,
\all  S_{j} \in \calM_{j}^{v}, \exists S_{j'} \in \calM_{j'}^{v} \text{ with } S_{j} \subset S_{j'} \tag{HM$^{\ast}$}
\end{align}
\label{def:HM}
\end{Definition}
\vspace{-2\baselineskip}

Thus, a valuation has hereditary maximizers if a greedy algorithm selecting the highest marginal value item at each step always maintains, regardless of its tie-breaking decisions, a set of maximum value among sets of the same size. 
Example~\ref{ex:noGSnoS1subsetS2}'s valuation $v$ \emph{does not} have HM: 
$\calM_{1}^{v} \!=\! \{\{a\}\}$ 
 but $\calM_{2}^{v} \!=\! \{\{b,c\}\}$.  
 
However, a few well-studied classes of valuation functions are HM as we show shortly. 
Multi-unit valuations are one of the most basic combinatorial valuations.
A multi-unit valuation $v$ treats all the items identically. Consequently, for any $j$, $\calM_{j}^{v}$ is the collection of all sets of size $j$ and thus $v$ trivially has HM. 

\vspace{-0.5\baselineskip}
\begin{Lemma}
$\!$A multi-unit valuation 
has hereditary maximizers. 
\label{stmt:MUHM}
\end{Lemma}
\vspace{-0.5\baselineskip}

%

\label{sec:GS}

A valuation is {\em gross substitutes}, a well-studied condition in assignment problems~\cite{BikhchandaniO06AscendingPVA,GulS99WalrasianEwGS},   
if raising prices on some items preserves the demand on  other items.  
\vspace{-0.5\baselineskip} 
\begin{Definition}
A valuation $v$ is {\em gross substitutes}  (GS) 
if for any price vectors\footnote{We compare price vectors $\vec{p}, \vec{p'} \!\!\in\! \reals^{n}$ component-wise: $\vec{p}' \!\geq\! \vec{p} \!\iff\! p_{j}' \geq p_{j} \all j =1..n$.} $\vec{p}' \geq \vec{p}$, and any $ A \in \demandSet{}(\vec{p})$ there exists $A' \in \demandSet{}(\vec{p}')$ with $ A' \supseteq \{i\in A: p_{i} = p'_{i} \}$.
\label{def:GS}
\end{Definition}

Remarkably~\cite{GulS99WalrasianEwGS}, for any set of GS buyers with public valuations, 
there exists a Walrasian (or competitive) equilibrium with one-shot item (possibly non-uniform) prices, i.e. at which buyers' preferred  bundles form a partition of all items. 
 Among GS valuation classes (see~\cite{LienY07OnTheGrossSC} for  more examples) are unit demand valuations (that define the value of a set as the highest value of an item within the set) and 
 concave multi-unit demand valuations. 
We know from  Lemma~\ref{stmt:MUHM} that the latter valuations have HM --   this is not a coincidence.  

\vspace{-0.5\baselineskip} 
\begin{Theorem}~\cite{Bertelsen04SubstitutesVaMC} A gross substitutes valuation has hereditary maximizers. 
\label{stmt:GSHM}
\end{Theorem}
\vspace{-0.5\baselineskip} 

Bertelsen~\cite{Bertelsen04SubstitutesVaMC} implicitly proves 
Theorem~\ref{stmt:GSHM}, without defining HM.  
Appendix~\ref{app:GSHM} provides a simpler proof for it 
via a basic graph-theoretic fact starting, like~\cite{Bertelsen04SubstitutesVaMC}, 
from Lien and Yan's~\cite{LienY07OnTheGrossSC} GS characterization (Lemma~\ref{stmt:GSabcS} in Appendix~\ref{app:GSHM}). 
The sets $\calM_{j}^{v}$ have more structure for a GS $v$.\footnote{\label{fn:GSabc}
\newcommand{\andF}{\text{ and }}
E.g. 
if ($\{a\} \in \calM_{1} \andF \{b,c\} \in \calM_{2}$) then ($\{b\} \in \calM_{1} \andF \{a,c\} \in \calM_{2}$) or ($\{c\} \in \calM_{1} \andF \{a,b\} \in \calM_{2}$)}
The high-level idea of  Theorem~\ref{stmt:GSHM}'s proof is as follows. We define a (bipartite) directed graph among certain sets of equal size; an edge from set $S$ to set $S'$ shows that $S$ has a strictly higher certain marginal value in $v$ than $S'$. 
If $v$ did not have HM, then this graph would have a directed cycle which is impossible. 


\newcommand{\pairBasedHMnGS}{
Assume that items are partitioned into pairs (e.g. movie--sequel) and any item is valued at $1$. 
The value of a set $S$ equals $|S|$ plus a function increasing in the number of such pairs in $S$.}

\newcommand{\singleMnHM}
{
each buyer $i$ has a desired bundle $S_{i}$ such that $v(S') = v(S_{i})$ if $S'$ contains $S_{i}$ and $v(S')=0$ otherwise. 
}

\label{sec:HMnGSEx}
We now exhibit a few valuation classes that are HM, but not GS (properties proved in Appendix~\ref{app:exHM}). They attest to the richness of our HM class, even when compared to the well-studied GS 
class. 
Intuitively, item complementarities  are severely limited by 
the GS property and to a lesser extent by the HM property.
\vspace{-0.5\baselineskip} 
\begin{description}
 \setlength{\itemsep}{1pt}
  \setlength{\parskip}{0pt}
  \setlength{\parsep}{0pt}
\item[order-consistent] 
For any $L \in 1..n$ and any sets $\{j_{1}, \dots, j_{L}\}$ and $\{j_{1}',\dots, j_{L}'\}$, whenever $v(\{j_{l}\}) \geq v(\{j_{l}'\}), \all l = 1..L$ then $v(\{j_{1}, \dots, j_{L}\}) \geq v(\{j_{1}', \dots, j_{L}'\})$, with strict inequality if at least one single item inequality is strict.  
\item[sequence-based] Assume that all items form a series (of e.g. episodes) 
and any item is valued at $1$. 
Then the value of a set $S$ equals $|S|$ plus a function increasing in the number of consecutive items in $S$. 
\item[pair-based] \pairBasedHMnGS
\end{description}
\vspace{-0.5\baselineskip} 


We proceed with a quantity guarantee for HM valuations, 
%
that will be critical 
for guarantees on sequential revenue. 
No fewer items are sold for price sequence $r^{1}, \dots, r^{d}$ (regardless of which preferred bundles are bought) than in  the worst-case for $r^{d}$ alone, i.e. $F(r^{d})$. 
This guarantee follows from a strong structural property, that we highlight for $d\!=\!2$.  
Any set $S_{1} \!\in\! \demandSet{}(r^{1})$ (i.e. preferred at a higher linear uniform price $r^{1}$) 
can serve as base to create sets preferred at the lower price $r^{2}<r^{1}$ 
via joining any set $S_{2} \!\in\! \demandSet{S_{1}}(r^{1},r^{2})$  (i.e. preferred sequentially at $r^{2}$ after buying $S_{1}$):  formally, 
 $S_{2} \cup S_{1} \!\in\! \demandSet{}(r^{2})$. 

\vspace{-1\baselineskip} 
 \begin{Theorem}
Fix an HM valuation $v$, a day $d \leq k$ and prices $r^{1} > \dots > r^{d}$. 
\noindent
Let $S_{\delta} \!\in\! \demandSet{S_{1}, \ldots, S_{\delta-1}}(r^{1}, \dots, r^{\delta})$ preferred at $r^{\delta}$ given sets $S_{1},\! \ldots,\! S_{\delta-1}$ sequentially bought at $r^{1}\!, \!\ldots\!, r^{\delta-1} \all \delta \!=\! 1..d$.  
Then $\bigcup_{\delta=1}^{d} \!S_{\delta} \!\in\! \demandSet{}(r^{d})$ and  thus
%
$\sum_{\delta=1}^{d} \!|S_{\delta}| \!\geq\! F(r^{d})\!$.
\label{stmt:oidDS}\end{Theorem} 
\vspace{-1\baselineskip} 



We first state a property needed in Theorem~\ref{stmt:oidDS}'s proof. 
 $\!$Clearly, a size $j$ set (if any) preferred at a uniform price cannot have a higher value than another size $j$ set. 
\begin{Lemma}
For all prices $r$ and sizes $j$, $\demandSet{}(r) \!\cap\! \{ |S| = j\}$ is either empty or $\calM_{j}^{v}.\!$ 
\label{stmt:demandSetM}
\end{Lemma}
\vspace{-0.5\baselineskip}

\begin{proof}[of Theorem~\ref{stmt:oidDS}]
We treat the case $d=2$; the claims for general $d$ follow similarly. 

Let $S_{1} \!\in\! \demandSet{}(r^{1})$  {be a set preferred at price $r^{1}$} and assume $|S_{1}| < F(r^{2})$ (otherwise the claim is immediate). Let $S_{2} \!\in\! \demandSet{S_{1}}(r^{1},r^{2})$  {be a set preferred at price $r^{2}$ after having bought $S_{1}$ at price $r^{1}$}. 
By Lemma~\ref{stmt:demandSetM}, $S_{1}\in \calM_{\emptyset,|S_{1}|}$. 
%
As $F(r^{2}) \!>\! |S_{1}|$, by (HM$^{\ast}$), $\exists S_{2}' \!\in\! \calM_{\emptyset,F(r^{2})}$  {a minimal set preferred at price $r^{2}$}
with $S_{1} \subset S_{2}'$. 
As $\calM_{\emptyset,F(r^{2})} \cap \demandSet{}(r^{2}) \neq \emptyset$  {(it contains ${S_{2}'}$)}, 
by Lemma~\ref{stmt:demandSetM}, 
 $S_{2}' \in \demandSet{}(r^{2})$.

 

\newcommand{\utSTwoMSOne}{u_{S_{2}'\setminus S_{1}}}

Let $u_{S} = v(S \cup S_{1}) - r^{1}|S_{1}| - r^{2}|S|$ be the utility from buying $S \subseteq 1..n\!\setminus\! S_{1}$ at $r^{2}$ after buying $S_{1}$ at $r^{1}$. 
As $S_{2} \!\in\! \demandSet{S_{1}}(r^{1},r^{2})$, $\utSTwoMSOne-u_{S_{2}} = (v(S_{2}') - r^{2}|S_{2}'|) - (v(S_{2} \cup S_{1}) - r^{2}|S_{2} \cup S_{1}|) \leq 0$. 
%
If $\utSTwoMSOne \!<\! u_{S_{2}}$ then $S_{2}\cup S_{1}$ is preferred to $S_{2}'$ at $r^{2}$, contradicting $S_{2}' \!\in\! \demandSet{}(r^{2})$.  Thus $\utSTwoMSOne\!=\!u_{S_{2}}$ implying 
$S_{2}\cup S_{1} \!\in\! \demandSet{}(r^{2})$. 
%
%
\qedF\end{proof}

We see that Theorem~\ref{stmt:oidDS} relies on the HM property: 
its conclusions fail 
for  Example~\ref{ex:noGSnoS1subsetS2}'s non-HM submodular valuation.
 A much weaker statement (proved in Appendix~\ref{app:submodS12}) than Theorem~\ref{stmt:oidDS} holds for such valuations. 
\vspace{-0.5\baselineskip} 
\begin{Lemma} 
For a \emph{submodular} valuation $v$, if $S_{1} \in \demandSet{}(r^{1}), S_{2}\in \demandSet{S_{1}}(r^{1}, r^{2})$ and $S_{2}' \in \demandSet{}(r^{2})$ 
then $S_{2} \not \supset S_{2}'\!\setminus\! S_{1}$ (note that equality is allowed). 
\label{stmt:submodS12}
\end{Lemma}










\vspace{-1\baselineskip}
\section{Revenue approximation for independent HM valuations}
\label{sec:HMApprox}
\vspace{-\baselineskip}

  \renewcommand{\P}{}

\newcommand{\LL}{L} %
\newcommand{\LLStart}{0} %
\newcommand{\LValue}{1+\log m + \log n}

We now leverage Theorem~\ref{stmt:oidDS}'s guarantees  to obtain a revenue approximation, complementing 
our $\frac{1+\log n}{2k} \!\leq\! \revApproxFactorMath{m}{n}{k}$ lower bound for $1\!+\!\log n \!\leq\! m$.  
Let $\LL = \LValue$.

\vspace{-0.5\baselineskip} 
\begin{Theorem}[upper bound]
$\!$Consider $m$ 
HM valuations with maxima $H_{1}\!\dots\!H_{m}$ and let $H \!=\! \max_{i\in \buyers} \!H_{i}\!$.  
Consider $k$ prices $q_{x_{1}} = \HS{x_{1}} \geq \dots \geq q_{x_{k}} = \HS{x_{k}}$ where  $x_{1} \leq \dots \leq x_{k}$ are the first (lowest), $\dots$, $k$-th (highest) order statistics of $k$ iid $U[\LLStart, \LL]$ continuous random variables  $u_{1}, \dots, u_{k}$. 
These prices (sorted decreasingly) yield expected revenue 
$\Omega(\frac{k}{\log m + \log n}) \!\sum_{i \in \buyers}\! H_{i}$. 
Thus $\revApproxFactorMath{m}{n}{k} \!=\! O(\!\frac{\log m + \log n}{k}\!)\!$.
\label{stmt:logn/k}
\end{Theorem}

If  valuations have the same maximum ($H_{i} = H, \all i \in\! \buyers$) 
then, as in~\cite{BalcanBM08ItemPfRM}, the approximation factor can be improved  to $\Omega(\frac{\log n}{k})$ 
  by using $\LL = \log (2n)$. 
 Recalling Theorem~\ref{stmt:lognExpScaled}'s lower bounds, 
we see that \emph{our  bounds are  tight} modulo a constant factor.


Before proceeding with Theorem~\ref{stmt:logn/k}'s proof, we  review, 
motivated by Lemma~\ref{stmt:Fdecrp}, a natural analogue of a well-studied economic concept and relate it  to $H$. 


A valuation $v$'s {\em demand curve}~\cite{BalcanBM08ItemPfRM} is a step function given by $(p_{l},F(p_{l}))_{l = 0..n_{v}+1}$ (with $n_{v} \!\leq\! n$) where threshold prices $0 = p_{0} \!<\! p_{1} \!<\! .. \!<\! p_{n_{v}} \!\leq \! p_{n_{v}+1} \!=\! H$  satisfy $F(p_{l}) = F(p) > F(p_{l+1}), \all p \in [p_{l}, p_{l+1}), \all l = 0..n_{v}$. 
That is, for any $l$ and any price $p$ in $[p_{l}, p_{l+1}]$ the lowest size of a (preferred)  bundle in $\demandSet{}_{v}(p)$ is $F(p_{l})$. 
The area $\areaF{F}$ under $v$'s demand curve is defined as $\sum_{l=1}^{n_{v}} p_{l} (F(p_{l}) - F(p_{l+1}))$. 
\begin{Lemma}\cite{BalcanBM08ItemPfRM}
 $\areaF{F} = H = \max_{S \subseteq 1..n} v(S)$, i.e. $v$'s maximum willingness to pay. 
\end{Lemma}
\vspace{-0.5\baselineskip}

The (worst-case) revenue $pF(p)$ from a single price $p$ equals the part of $\areaF{F}$ covered by $p$. 
We now generalize this to a sequence of prices. 
 For instance, prices $r^{1} \!>\! r^{2}$ cover a 
$F(r^{1}) r^{1} + (F(r^{2})\!-\!F(r^{1}))r^{2}$ part of $\areaF{F}$, i.e. the area of the union of two 
rectangles with opposite corners $(0,0)$ and $(r^{i}, F(r^{i})).\!$ 
Note that no pricing scheme can cover more than $\areaF{F}$ itself. 
 However, as seen in Example~\ref{ex:noGSnoS1subsetS2}, the area covered by a scheme with at least two prices may  be less than its revenue even for a submodular valuation. 

\vspace{-0.5\baselineskip} 
\begin{Definition}
The fraction of $\areaF{F}$ 
\emph{covered} by a pricing scheme $\calP$ 
with prices $p'_{1} \!>\! \dots \!>\! p'_{k}$ is \\
$(1/\areaF{F}) \sum_{d=1}^{k} p'_{d} (F(p'_{d}) - F(p'_{d-1}))$ where $F(p'_{0})=0$.
\label{def:PrCover}
\end{Definition}
\vspace{-0.5\baselineskip}

We are now ready for Theorem~\ref{stmt:logn/k}'s proof. 
It establishes that for HM valuations the revenue of a pricing scheme $\calP$ is at least the part of $\areaF{F}$ 
covered by $\calP$. 
%

\begin{proof}[of Theorem~\ref{stmt:logn/k}] 
%
We proceed with one buyer; linearity of expectation will then yield the claim. 
Let $q_{x_{1}} \!\geq \!\dots\! \geq\! q_{x_{k}}$ be the prices of Theorem~\ref{stmt:logn/k} for $H$.  
Let set $S_{d}' \! \in\! \demandSet{S'_{1},\! \dots\!, S'_{d-1}}(q_{x_{1}}\dots q_{x_{d}} )$ be  bought in day $d$.  
By Theorem~\ref{stmt:oidDS}, 
$\sum_{\delta=1}^{d} |S'_{\delta}| \!\geq\! F(q_{x_{d}})$. 

Via Lemma~\ref{stmt:groupSold} with 
 $d_{0}\!=\!1,d\!=\!k, q^{\delta}\!=\!q_{x_{\delta}}$ and $x^{\delta}\!=\!F(q_{x_{\delta}})$, revenue is at least  
 $\sum_{\delta=1}^{d} q_{x_{\delta}} (F(q_{x_{\delta}}) - F(q_{x_{\delta-1}}))$, i.e. the area covered by these prices.  
 Theorem~\ref{stmt:coverk/logn} will yield the approximation factor. 
\qedF \end{proof}

We have reached the technical core of our approximation: the $k$ random\footnote{While Balcan et al.~\cite{BalcanBM08ItemPfRM} 
also use prices of the form $q_{x} = \HS{x}$ to achieve the $\Theta(\log m + \log n)$ factor for $k=1$ day, 
we find it more convenient to use continuous $x$'s (scaling factors) as opposed to integer. 
} prices cover well in expectation the area under the demand curve of a valuation.  
The area covered by the $k$ prices is the sum of the areas of each individual price minus the area  covered by each of at least two prices. 
Theorem~\ref{stmt:coverk/logn}'s proof (in Appendix~\ref{app:HMApproxProofs}) uses properties of order statistics to upper bound the latter area. We note that 
Akhlaghpour et al.~\cite{AkhlaghpourGHMMN09OptimalIPwPNE} present a recursive construction that can be used to 
cover in expectation 
an $\Omega(\frac{\log k}{\log m + \log n})$ fraction of $\areaF{F}$   
as opposed to our $\Omega(\frac{k}{\log m + \log n})$ factor. 

\newcommand{\coverklognStmt}{
Consider $k$ prices $q_{x_{1}} \!=\! \HS{x_{1}} \!\geq\! \ldots \!\geq\! q_{x_{k}} = \HS{x_{k}}$ where $x_{1} \leq \dots \leq x_{k}$ are the first (smallest), $\dots$, $k$-th (largest) order statistics of $k$ iid continuous random variables $u_{1}, \dots, u_{k}$ chosen uniformly at random in $[\LLStart, \LL]$. 
Then these prices 
cover in expectation 
 an $\Omega(\frac{k}{\log m + \log n})$ fraction of $\areaF{F}$. 
}

\begin{Theorem}
\label{stmt:coverk/logn} 
\coverklognStmt
\end{Theorem}


We finally provide a revenue guarantee given a guarantee on total quantities bought. 
\vspace{-0.5\baselineskip}
\begin{Lemma}
If, at prices $q^{d_{0}} \!>\! \ldots \!>\! q^{d}$, 
at least $x^{\delta}$ items ($x^{d} \!\geq\! x^{d-1} \!\geq\! \ldots \!\geq\! x^{d_{0}} \!\geq\! x^{d_{0}-1}=0$) are sold in total up to each day $\delta = d_{0}..d$ 
(e.g. at least $x^{d_{0}+1}$ items in days $d_{0}$ and $d_{0}+1$ together)
then the revenue is at least 
$\tst\sum_{\delta=d_{0}}^{d} q^{\delta} (x^{\delta} - x^{\delta-1})$. 
\label{stmt:groupSold}
\end{Lemma}
\vspace{-0.5\baselineskip}
\begin{proof} \ 
The lowest revenue is for \emph{exactly} $x^{\delta}$ items sold in day $\delta = d_{0}..d$ and (as prices are decreasing) 
for as few items as possible sold early, i.e. for 
$x^{\delta} \!-\! x^{\delta-1}$ items sold in day $\delta \!=\! d_{0}..d$.\qedF 
\end{proof}


 In practice, buyers do not only have patience, but also have an influence on other buyers. We allow now a buyer's value (for any bundle) to increase depending on others' acquired  bundles (but not on others' valuations). We will preserve the revenue approximation factor, despite an increased optimum revenue.

\section{Positive allocative externalities}
\label{sec:extys}
\vspace{-0.5\baselineskip}

We now investigate revenue maximization in the presence of positive externalities, i.e. a buyer's valuation being increased by other buyers' ownership of certain items. Such influences can be subjective, e.g. resulting from peer pressure, or objective, e.g. resulting from ownership of a certain social network application. 


\newcommand{\inflProp}{\ensuremath{\mathcal{I}}}

\noindent 
We define a new influence model via a predicate $\inflProp\!:\!1..m\!\to\!\{\texttt{false},\texttt{true}\}$ such that $\inflProp(i_{0})$ only depends on seller's assignment of items to buyer $i_{0}$, e.g. 
\vspace{-0.5\baselineskip}
\begin{itemize}
 \setlength{\itemsep}{1pt}
  \setlength{\parskip}{0pt}
  \setlength{\parsep}{0pt}
\item $\inflProp(i_{0}) = \texttt{true}$ iff buyer $i_{0}$ owns all (or, instead, at least two)  items
\item $\inflProp(i_{0}) = \texttt{true}$ iff buyer $i_{0}$ owns his preferred bundle at current prices
\end{itemize}
\vspace{-0.5\baselineskip}
Let $\inflProp_d$ be the buyers $i_{0}$ satisfying $\inflProp(i_{0})$ before day $d$. $\inflProp$ is \emph{monotone} if 
$\inflProp_{d} \subseteq \inflProp_{d+1}$. 

We model the valuation in day $d$ of a buyer $i$ as a linear mapping (depending on $d$ only through its argument $\inflProp_{d} \wo{i}$) 
of ${i}$'s  base value
\vspace{-0.5\baselineskip} 
\begin{align}
v_{i}^{d} (S | \buyers\wo{i}) = (a_{i}(\inflProp_d \wo{i}) v_{i}(S)) \oplus b_{i}(\inflProp_d \wo{i}), \all \: \text{set } S \subseteq 1..n
\label{eq:linExtys}
\end{align}
%
where 
$\alpha v_{i}(S) \oplus \beta = \{\alpha v_{i}(S) \!\text{ if }\! S \!=\! \emptyset$ and\footnote
{ {$\!$Eq.~\eqref{eq:linExtys} excludes the additive increase for $S\!=\!\emptyset$ 
so that $v_{i}^{d}(\emptyset|\cdot) \!=\! 0$. 
Also, if 
$b_{i}$ is much larger than $a_{i} \max_{S} v_{i}(S)$ then 
 a multiplicative revenue approximation is impossible: prices close to $b_{i}$ are needed, rendering $\emptyset$ the preferred set, i.e. zero revenue. 
}}
$\alpha v_{i}(S)+\beta \!\text{ if }\! S \!\neq\! \emptyset\}$ for  $\alpha, \beta \in \reals$. 

Thus, $a_{i}(I)$ and $b_{i}(I)$ measure the multiplicative and additive 
influences that a buyer set $I$ (satisfying \inflProp) have on buyer $i$. Say $i$'s value for any DVD of a TV series doubles as soon as one other friend (in a set $F_{i}$)  has  the entire series (the predicate $\inflProp$)
and is then constant. Then $a_{i}(I) = 2 \!\!\iff\!\! |I \cap F_{i}| \geq 1$ and $b_{i}(I) = 0$.  

Without any influence, a valuation reduces to the base value: $a_{i}(\emptyset) = 1, b_{i}(\emptyset) = 0$. 
Assume $a_{i}$ and $b_{i}$ are non-negative, monotone and submodular\footnote{Submodularity (non-increasing marginal influence) is often  assumed for externalities~\cite{AkhlaghpourGHMMN09OptimalIPwPNE,HartlineMS08OptimalMSoSN}. Positive, monotone externalities are an instance of ``herd mentality''.}. 
Also assume that  $a_{i}, b_{i}, v_{i}$ are bounded: $\max_{I \subseteq \buyers\wo{i}} a_{i}(I) = a_{i}(\buyers\wo{i})  = H^{a}, \ 
  \max_{I} b_{i}(I) = b_{i}(\buyers\wo{i})  = H^{b}, \ 
  \max_{S \subseteq 1..n} v_{i} (S) = H_{i}$ with $\max_{i\in\buyers} H_{i}=H$.  

Our influence model is a distribution-free extension of single-item models~\cite{AkhlaghpourGHMMN09OptimalIPwPNE,HartlineMS08OptimalMSoSN}.  
It does not require or preclude symmetry, anonymity or a neighbor graph. 
For $a_{i} = 1, b_{i} = 0$ we recover the model before this section. Buyers are still forward-myopic and do not strategize about which items to buy today 
so that other buyers' values increase, thus increasing their own value etc.

\newcommand{\inflADa}[2]{\varphi_{a}^{#1}(#2)}
\newcommand{\inflADb}[2]{\varphi_{b}^{#1}(#2)}

\noindent
With positive externalities, a natural revenue maximization approach~\cite{HartlineMS08OptimalMSoSN} is  providing certain items for free to some buyers and then charging others accordingly. 
\vspace{-0.5\baselineskip} 
\begin{Definition}  
The influence-and-exploit $IE_{k}$ marketing strategy for $k \geq 2$ 
satisfies $\inflProp$ (at no cost) for each buyer with probability $0.5$, in day $1$. 
Let $A_{1}$ be the set of buyers chosen in day $1$. 
Independently of $A_{1}$, 
 $k-1$ prices $q_{x_{1}} \!=\! \HSEx{x_{1}} \!\geq\! \ldots \!\geq\! q_{x_{k-1}} = \HSEx{x_{k-1}}$ where $x_{1} \leq \dots \leq x_{k-1}$ are the first (smallest), $\dots$, $(k\!-\!1)$-th (largest) order statistics of $k\!-\!1$ iid continuous random variables $u_{1}, \!\dots\!, u_{k-1}$ chosen uniformly at random in $[\LLStart, \LL]$. 
Each buyer $i \!\in\! \buyers \!\setminus\!A_{1}$
is offered uniform item price $\scfa{i} \cdot q_{x_{d-1}}\!$ in day $d \!\geq\! 2$.
\label{def:IEn} 
\end{Definition}
\vspace{-0.5\baselineskip}

\noindent
This section's main result (proved in Appendix~\ref{app:extys}), is 
that Theorem~\ref{stmt:logn/k}'s factor   
carries over to externalities, despite the affine increase in the optimum revenue.
%
\vspace{-0.5\baselineskip} 
\begin{Theorem}
$\!\!$The $IE_{k}\!$ strategy is an $O(\frac{\log m + \log n}{k})$-revenue  approximation to the optimal marketing strategy for a monotone $\inflProp$ over $IE_{k}$ and HM base valuations. 
\label{stmt:logn/kExtys}
\end{Theorem}
%

The price schedule 
$q_{x_{1}} \!\geq\! \ldots \!\geq\! q_{x_{k-1}}$
is (by Theorem~\ref{stmt:logn/k}) 
 a $O(\frac{\log m + \log n}{k})$-revenue approximation given buyers' base valuations (translated by  $\frac{\scfb{i}}{\scfa{i}}$). 
The proof establishes that 
 the influence of other buyers (an affine mapping of a buyer's value in each day) does not result in fewer items being bought in the worst case. 
%

\section{Conclusions and future directions}
\vspace{-0.5\baselineskip}

In this paper we study prior-free revenue maximization with sequences of equal item prices.  
We are the first to consider combinatorial valuations for more than one item in unlimited supply in the sequential setting. We provide a sufficient condition and an algorithm 
 improving  the revenue approximation factor of an existing one-shot pricing scheme complemented by a lower bound that leverages the limited availability of price updates. 
Our paper also initiates the study of revenue maximization for allocative externalities between combinatorial valuations. For  positive, non-anonymous externalities we present a simple marketing strategy preserving the approximation factor without externalities, despite an increase in revenue available due to such influences. Several open directions appear promising  to us.


Hereditary maximizers guarantees consistency of bundles bought sequentially. We deem it of interest to find an alternative assumption,  
perhaps related to sequential revenue instead, that still allows revenue bounds.  
%
%
 {$\!$We assume fully patient, as opposed to instantaneous, buyers. $\!$Other patience models, e.g. $\!$arrival-departure intervals~\cite{BansalCCRSS10DynamicPfIB}, may yield alternative approximations.}
%
%
Finally, widespread externalities in applications present many exciting open questions, both practical and theoretical, notably in multiple-item settings.  

\noindent
\textbf{Acknowledgments}. We thank Avrim Blum and Malvika Rao for detailed comments on earlier drafts of this paper, 
Mark Braverman 
for helpful discussions and Daniel Lehmann for providing us with a copy of~\cite{Bertelsen04SubstitutesVaMC}.

\vspace{-\baselineskip}

\bibliographystyle{abbrv}
\bibliography{seqlPrUnlSupplySAGT}
 
 \appendix
 
%
%



 \section{Alternate proof of Theorem~\ref{stmt:GSHM}}
 \label{app:GSHM}
 
 Denote $v$'s \emph{conditioning} on set $S$ (measuring marginal value over $S$) by $
v^{S}(A) = v(S \cup A) - v(S), \all A \subseteq 1..n \!\setminus\! S
$. 

As mentioned in Section~\ref{sec:HM}, 
there is a more direct, valuation-based (as opposed to price-based as in Def.~\ref{def:GS}) characterization of GS valuations. 
\begin{Lemma}~\cite{LienY07OnTheGrossSC}
 $v$ is gross substitutes if and only if 
$v$ is submodular
and 
\begin{align}
\label{eq:GSabcS}
\text{\hspace{-0.2in} $\all$ items $a,b,c,\!\!$ set } S, 
v^{S}(ab)+v^{S}(c) \leq \max\{v^{S}(ac)+v^{S}(b), v^{S}(bc)+v^{S}(a)\}
\end{align}
\label{stmt:GSabcS}
i.e. no unique maximizer among $v^{S}(ab)+v^{S}(c), v^{S}(ac)+v^{S}(b), v^{S}(bc)+v^{S}(a)$. 
\end{Lemma}
For $S\!=\!\emptyset$, this immediately leads to the observation in Footnote~\ref{fn:GSabc} in Section~\ref{sec:HM}.

\newcommand{\mvcap}[1]{v_{\!\scriptscriptstyle\cap\!}^{#1}} 
\newcommand{\vmcap}[1]{\mvcap{#1}}

We now prove a slightly stronger result than Theorem~\ref{stmt:GSHM}'s claim in Section~\ref{sec:HM}.
\smallskip

\noindent
\textbf{Theorem~\ref{stmt:GSHM} } 
A GS valuation $v$ has hereditary maximizers fixing any base bundle 
already owned.

 \newcommand{\bipGraphName}{\ensuremath{G_{L-1,c_{1}L-2}}}
\begin{proof}
Suppose towards a contradiction that $v$ did not have HM: i.e. for some 
 $S_{j} \in \calM_{j}$ no $S_{j+1} \in \calM_{j+1}$ contains it. 
 Choose  $S_{j+1}$ with lowest $|S_{j} \!\setminus\! S_{j+1}| + |S_{j+1} \!\setminus\! S_{j}|$.   

 For any set $S$, let $\mvcap{B}(S) = v^{(S_{j}\cap S_{j+1})\cup B}(S)$ (simply $\vmcap{}$ if $B=\emptyset$).

Assume that 
\begin{align}
\tst\max_{x \in S_{j} \!\setminus\! S_{j+1}} \mvcap{}(\{x\}) \geq 
\tst\max_{x \in S_{j+1} \!\setminus\! S_{j}} \mvcap{}(\{x\})
\label{eq:vmcapBetterS1-S2}
\end{align} 
and let $c_{1} \in \argmax_{x \in S_{j} \!\setminus\! S_{j+1}} \mvcap{}(\{x\})$.

We prove inductively that for any $L$ distinct items  $a^{1}...a^{L} \in S_{j+1}\!\setminus\!S_{j}$,
\begin{align}
\exists l \in 1..L \text{ with } \vmcap{}(a^{1},\dots,a^{L}) \leq \vmcap{}(a^{-l},c_{1}) 
\label{eq:GSIndHyp}
\end{align}
where $a^{-l}$ denotes $a^{1} ... a^{l-1} a^{l+1} ... a^{L}$. 
The theorem follows for $L_{21} = |S_{j+1}\!\setminus\!S_{j}|$: $\vmcap{}(S)$ is maximized by $S_{j+1}\!\setminus\!S_{j}$ among sets of size $L_{21}$. By Eq.~\eqref{eq:GSIndHyp}, $(S_{j+1}\wo{a^{l}})\wi{c_{1}} \in \calM_{j+1}$ and  has strictly fewer elements in the symmetric difference with $S_{j}$ than $S_{j+1}$, contradicting the choice of sets $S_{j}$ and $S_{j+1}$.

As a base case $L=1$, Eq.~\eqref{eq:GSIndHyp} holds by choice of $c_{1}$ as $l=1$.

Assume that Eq.~\eqref{eq:GSIndHyp} holds for $L-1$, 
and suppose  it fails for $L \geq 2$. 

We define a directed bipartite graph \bipGraphName\ with vertices  $a^{-l}$
for $l =1..L$ in one partition (that we call $\calP_{L-1}$) and $c_{1}a^{-h,l}$
for $1\leq l < h \leq L$ in the other partition (that we call $\calP_{c_{1}L-2}$). 
Directed edge $a\to b$ in \bipGraphName\ represents $\vmcap{}(a) \leq \vmcap{}(b)$, with strict inequality if $a \in \calP_{c_{1}L-2}$ and $b \in \calP_{L-1}$.

We show that in \bipGraphName\ each vertex has at least one outgoing edge, i.e. there exists a cycle of inequalities (at least half of them strict),  contradiction. This claim holds for vertices in $\calP_{L-1}$ by the inductive hypothesis. 

Fix $1\leq h < l\leq L$ and $c_{1}a^{-h,l}$.
The failure of Eq.~\eqref{eq:GSIndHyp} for $L$ requires  
\begin{align}
\vmcap{}(a^{1}\!,...,a^{L}) \!&>\!
 \max\{\vmcap{}(a^{-l}\!,c_{1}), \vmcap{}(a^{-h}\!,c_{1}) \} \text{ i.e. } \label{eq:c1-hlA}  \\
\vmcap{a^{-h,l}}\!(a^{h}\!,a^{l}) \!&>\!
 \max\{\vmcap{a^{-h,l}}\!(a^{h}\!,c_{1}), \vmcap{a^{-h,l}}\!(a^{l}\!,c_{1}) \}\text{ implying, via Eq.~\eqref{eq:GSabcS}}\! \label{eq:c1-hlB} \\
\vmcap{a^{-h,l}}\!(c_{1}) \!&<\!
 \max\{\vmcap{a^{-h,l}}\!(a^{h}), \vmcap{a^{-h,l}}\!(a^{l})\} \text{ i.e. } \label{eq:c1-hlC} \\
\vmcap{}(c_{1}a^{-h,l}) \!&<\!
 \max\{\vmcap{}(a^{-l}), \vmcap{}(a^{-h}) \} \label{eq:c1-hlD}
\end{align}
 exhibiting one outgoing edge from $c_{1}a^{-h,l}$, i.e. to $a^{-l}$  or $a^{-h}$. 
 
 If Eq.~\eqref{eq:vmcapBetterS1-S2} did not hold, then one can show as above, 
 that for any $L$ distinct items  $b^{1}...b^{L} \in S_{j}\!\setminus\!S_{j+1}$,
and $c_{2} \in \argmax_{x \in S_{j+1} \!\setminus\! S_{j}} \mvcap{}(\{x\})$
\begin{align}
\exists l \in 1..L \text{ with } \vmcap{}(b^{1},\dots,b^{L}) < \vmcap{}(b^{-l},c_{2}) 
\label{eq:GSIndHypRev}
\end{align}
 For $L = |S_{j}\!\setminus\!S_{j+1}|$, this contradicts 
$S_{j} \in \calM_{j}$. 
\end{proof}

\begin{figure}[!b]
  \begin{center}
   \includegraphics[width=0.5\textwidth]{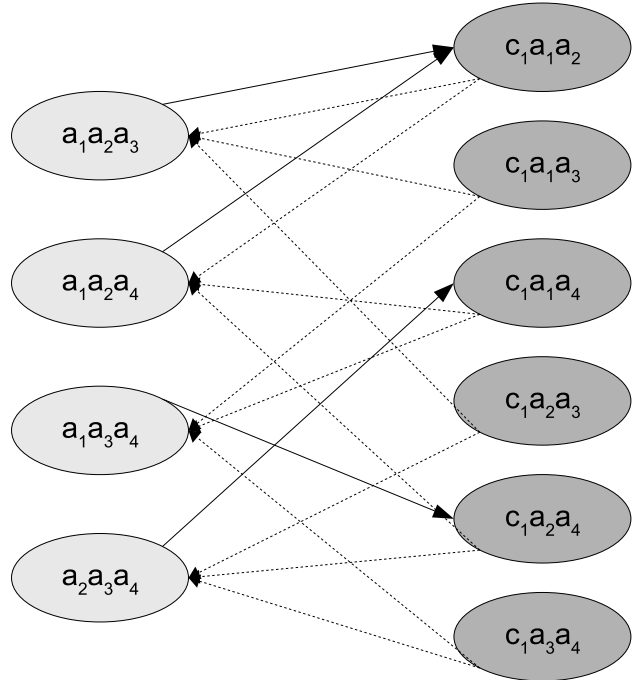}
\vspace{-2\baselineskip}
  \end{center}
  \caption{Illustration of Theorem~\ref{stmt:GSHM}'s alternate proof for $L\!=\!4$. Graph $G_{3,c_{1}2}$ has left-hand side partition $\calP_{3}$ 
  and right-hand side partition $\calP_{c_{1}2}$.  
  Any edge from set $S$ to set $S'$ encodes that $S$ has a lower marginal value than $S'$ over $S_{j} \cap S_{j+1}$ where $S_{j}$ and $S_{j+1}$ are  value-maximizing sets of size $j$ and $j+1$ with minimum symmetric difference i.e. lowest $|S_{j} \!\setminus\! S_{j+1}| + |S_{j+1} \!\setminus\! S_{j}|$.  Theorem~\ref{stmt:GSHM}'s alternate proof  supposes  $S_{j} \not \subset S_{j+1}$ and obtains a contradiction via a cycle in $G_{3,c_{1}2}$, using 
elements $a_{1}, a_{2}, a_{3}, a_{4}, c_{1}$ not contained in $S_{j+1} \!\setminus\! S_{j}$. 
 $\calP_{3}$ contains one vertex for each set of $3$ elements in $S_{j+1} \!\setminus\! S_{j}$ whereas $\calP_{c_{1}2}$ 
  contains one vertex for each set comprised of $c_{1}$ and $2$ elements in $S_{j+1} \!\setminus\! S_{j}$. 
\newline  
By the induction hypothesis Eq.~\eqref{eq:GSIndHyp}, there exists an edge (solid in the figure) from any vertex on the left to some vertex on the right. 
Eqs.~(\ref{eq:c1-hlA}-\ref{eq:c1-hlD}) establish that there is an outgoing edge (dashed in the figure) from $c_{1}a^{-h,l}$ to $a^{-h}$  or $a^{-l}$, for $1 \leq l < h\leq 4$. 
 }
  \label{fig:GSHM}
  \vspace{-\baselineskip}
  \end{figure}

Bertelsen provides a stronger result: for a GS valuation $v$, the collection of sets 
$\calM_{j}^{v}$ for all $j$ is a \emph{greedoid},  
i.e. a collection $\calF$ of subsets of $1..n$ that is accessible (i.e. $\all S \in \calF, \exists x \in S $ with $S\wo{x} \in \calF$)
and satisfies the augmentation property 
($\all S, S' \in \calF$ with $|S|<|S'|$, $\exists x \in S'\!\setminus\!S $ with $S\wi{x} \in \calF$). 
Clearly, the maximum of an HM valuation without ties can be found efficiently via a  greedy algorithm -- Bertelsen provides such an algorithm  for a GS valuation that can also handle ties. 
\section{Examples of HM, but not GS, valuations}
\label{app:exHM}

\begin{description}
\setlength{\itemsep}{1pt}
\item[order-consistent] 
For any $L \in 1..n$ and any sets $\{j_{1}, \dots, j_{L}\}$ and $\{j_{1}',\dots, j_{L}'\}$, whenever $v(\{j_{l}\}) \geq v(\{j_{l}'\}), \all l = 1..L$ then $v(\{j_{1}, \dots, j_{L}\}) \geq v(\{j_{1}', \dots, j_{L}'\})$, with strict inequality if any of the single item inequalities is strict. 

Order-consistent valuations are HM because the maximum value among size $L$ valuations is attained at exactly those sets with item values among $L$ highest: say $k=3$, $v(1)=3, v(2)=v(3)=1$. Then an order-consistent valuation must have $v(1,2)=v(1,3)>v(2,3)$. This  valuation is not GS as Eq.~\eqref{eq:GSabcS} can be violated by setting $v(1,2)=v(1,3)=2, v(2,3)=1$. 

\item[sequence-based] Assume that all items form a series (of e.g. episodes) 
and that the value of any item is $1$. 
Then the valuation of a set $S$ equals $|S|$ plus a function that is increasing in the number of consecutive items in $S$. 

Sequence-based valuations are HM because the maximum value size $L$ sets coincide with length $L$ sequences. They are not GS as $v(1)+v(3,4) > \max\{v(1,3) +v(4), v(1,4)+v(3)\}$ contradicting Eq.~\eqref{eq:GSabcS}. 
%

\item[pair-based] \pairBasedHMnGS 

Pair-based valuations are HM because the maximum value size $L$ sets coincide with sets with $\lfloor \frac{L}{2} \rfloor$ pairs. They are not HM as 
$v(1)+v(3,4) > \max\{v(1,3) +v(4), v(1,4)+v(3)\}$ if $\{3,4\}$ is a pair, contradicting Eq.~\eqref{eq:GSabcS}. 
\end{description}
We note that HM only contains single-minded valuations if the bundle of interest is of size $1$: 
 \singleMnHM
Any bundle of size $|S_{i}|-1$ has value $0$, but only those included in $S_{i}$ can be augmented with one item to $S_{i}$.

\section{Proof 
of Lemma~\ref{stmt:submodS12}}
\label{app:oidDS}
\label{app:submodS12}

\begin{proof}[of Lemma~\ref{stmt:submodS12}]
By optimality of $S_{2}$ after having bought $S_{1}$, 
\begin{align} 
v(S_{1} \cup S_{2}) - r^{1}|S_{1}| - r^{2}|S_{2}| &\geq v(S_{1} \cup S) - r^{1}|S_{1}| - r^{2}|S|, \all S \subseteq \{1..n\}\!\setminus\! S_{1} \notag \\
%
%
\text{i.e. } v(S_{1} \cup S_{2}) - r^{2}|S_{2}|  &\geq v(S_{1} \cup (S_{2}' \!\setminus\! S_{1})) - 
r^{2} |S_{2}' \!\setminus\! S_{1}| \text{ for $S = S_{2}' \!\setminus\! S_{1}$}
\notag 
 \end{align}
 
 By moving $r^{2}|S_{2}' \!\setminus\! S_{1}|$ to the left-hand side and subtracting $r^{2}|S_{2}'|$ we get,   
\begin{align}
v(S_{1} \cup S_{2}) - r^{2} (|S_{2}| - |S_{2}' \!\setminus\! S_{1}| + |S_{2}'|) \geq v(S_{1} \cup S_{2}') - r^{2}|S_{2}'|. 
\label{eq:S12S2-S1}
 \end{align}
 
By optimality of $S_{2}'$ for price $r^{2}$ and $S_{2} \cap (S_{2}' \cap S_{1}) = \emptyset$ (as $S_{2}\cap S_{1} = \emptyset$),   
\begin{align}
v(S_{2}') - r^{2} |S_{2}'| 
\geq 
v(S_{2} \cup (S_{2}' \cap S_{1})) - r^{2} (|S_{2}| + |S_{2}' \cap S_{1}|) 
\label{eq:S2capS1}
 \end{align}
 
\noindent Adding Eqs.~\eqref{eq:S12S2-S1}~and~\eqref{eq:S2capS1} 
and canceling $r^{2}$ terms ($|S_{2}'| \!-\! |S_{2}' \!\setminus\! S_{1}| \!=\! |S_{2}' \cap S_{1}|$), 
\begin{align}
v(S_{1} \cup S_{2}) + v(S_{2}') &\!>\! v(S_{1} \cup S_{2}') + v(S_{2} \cup (S_{2}' \!\cap\! S_{1})) \text{ i.e. } \label{eq:S12submodS2Plus} \\
v(S_{2} \cup S_{1}) - v(S_{2} \cup (S_{2}' \!\cap\! S_{1})) &\!>\! v((S_{2}' \!\setminus\! S_{1})\cup S_{1}) 
- v(S_{2}')
\label{eq:S12submodS2Minus}
\end{align}
$S_{2} \not \supset S_{2}'\!\setminus\! S_{1}$  
 follows from $(S_{2}' \!\setminus\! S_{1}) \cup (S_{2}' \cap S_{1}) = S_{2}'$ and $v$'s submodularity$\!.\!$ 
\end{proof}




\newcommand{\calH}{\ensuremath{\mathcal{H}}}
\newcommand{\margValH}[1]{\ensuremath{v_{\scriptscriptstyle \calH}^{#1}}}

\vspace{-0.5\baselineskip}
\section{Proofs from Section~\ref{sec:HMApprox}}
\label{app:HMApproxProofs}
\vspace{-0.5\baselineskip}

Fix a valuation with maximum $H$ with area $\areaF{F}$ under its demand curve $F\!$. 
Let $L \!=\! \LValue$ and $AC_{1} \!=\! \expec_{l\sim U[0,L]} [F(q_{l}) q_{l}]$  
be the part of $\areaF{F}$ 
covered by a price $q_{l} = \HS{l}$ with $l$ distributed uniformly on $\LLStart..L$. 
 From~\cite{BalcanBM08ItemPfRM}, we know  
  $AC_{1} \!=\! \int_{0}^{L}  F(\frac{H}{2^{x}}) \frac{H}{2^{x}} L^{-1} \intd{x} \geq L^{-1}\sum_{x=1}^{L}  F(\HS{x}) \HS{x} \geq  \frac{H}{4(\LValue)}$ (note $F(H)\!=\!0$).

\textbf{Theorem~\ref{stmt:coverk/logn}} \  
\coverklognStmt
\smallskip  
  
  The following standard (see e.g.~\cite{DavidN03OrderS}, Chapter~2)  properties of order statistics will be needed in our Theorem's proof. 
  \begin{Lemma}
Let $X$ be a \emph{continuous} random variable in $[0,L]$ with cumulative distribution function  $F$ and probability density function $\!f \!=\! F'\!$. 
Let $\os{d}{k}$ 
be its $d^{\textrm{th}}\!$ highest order statistic out of $k$ independent trials. Then 
%
\begin{itemize}
\item $\os{j}{k}$'s probability density function is $k\binom{k-1}{j-1} F(x)^{j-1}(1-F(x))^{k-j} f(x)$. Thus $x_{d} = \os{d}{k}$'s ($F(x)=x/L, f(x)=1/L$)
probability density function is $k\binom{k-1}{d-1} x^{d-1}(L-x)^{k-d}L^{-k}$. 
\item $\os{d+1}{k}$'s distribution \emph{conditioned} on the next lowest $\os{d}{k}$'s value $x_{d}$ is the same as the distribution of the 
lowest order statistic $\os{k-d}{k-d}$ out of $k-d$ trials of variable $X$ truncated below $x_{d}$, i.e. with probability density function $\frac{f(x)}{1-F(x_{d})}$ for $x\in [x_{d}, L]$. 
\end{itemize}
\label{lemma:os}
  \end{Lemma}

%

%
\begin{proof} [of Theorem~\ref{stmt:coverk/logn}]
The expected area covered (recall Def.~\ref{def:PrCover}) by these prices is  
\begin{align}
AC_{k} &= \expec_{x_{1} \dots x_{k}} [\sum_{d=1}^{k} 
(F(q_{x_{d}}) - F(q_{x_{d-1}})) q_{x_{d}}] = 
\expec_{x_{1} \dots x_{k}} [\sum_{d=1}^{k} 
F(q_{x_{d}}) q_{x_{d}}]
- 
\expec_{x_{1} \dots x_{k}} [\sum_{d=1}^{k-1} 
F(q_{x_{d}}) q_{x_{d+1}}]  \notag \\
&= \expec_{u_{1} \dots u_{k}} [\sum_{d=1}^{k} F(q_{u_{d}}) q_{u_{d}}] \label{eq:orderStatsAreVarsk}
- \sum_{d=1}^{k-1} S_{k}^{d}
\quad 
= k AC_{1} - \sum_{d=1}^{k-1} S_{k}^{d} 
\end{align}
where Eq.~\eqref{eq:orderStatsAreVarsk} follows from the fact that the sets $\{x_{1}, \dots, x_{k}\}$ and  $\{u_{1}, \dots, u_{k}\}$ coincide 
and 
we denoted $S_{k}^{d} = \expec_{x_{1}\dots x_{k}} [F(q_{x_{d}}) q_{x_{d+1}}]$. 
We continue by upper bounding each $S_{k}^{d}$ and then summing them up.

Using Lemma~\ref{lemma:os} (the first and the second facts for Eq.~\eqref{eq:kRandomC1} and the first fact for Eq.~\eqref{eq:kRandomC3}), 
\begin{align}
\expec_{x_{d+1}|x_{d}} [q_{x_{d+1}}] &=
\tst\int_{x_{d}}^{L} \frac{(k-d)(L-y)^{k-d-1}}{(L-x_{d})^{k-d}} \frac{H}{2^{y}} \intd{y}
\leq 
\tst\int_{x_{d}}^{L} \frac{k-d}{L-x_{d}} \frac{H}{2^{y}} \intd{y} \leq 2 \frac{k-d}{L-x_{d}} \frac{H}{2^{x_{d}}}
\label{eq:kRandomC1}
\\ 
\text{Thus } S_{k}^{d} &=
\expec_{x_{1}\dots x_{k}} [F(q_{x_{d}}) q_{x_{d+1}}] 
= \expec_{x_{d}} [ F(q_{x_{d}}) \cdot  \expec_{x_{d+1}|x_{d}} [q_{x_{d+1}}] ]
\label{eq:kRandomC2} 
\\ 
&\leq \tst\int_{0}^{L} 
k \binom{k-1}{d-1} x_{d}^{d-1} (L-x_{d})^{k-d} L^{-k} F(\frac{H}{2^{x_{d}}})
 2 \frac{k-d}{L-x_{d}} \frac{H}{2^{x_{d}}} \intd{x_{d}}
 \label{eq:kRandomC3}
  \\ 
 &\leq \tst 2 k(k-1) \binom{k-2}{d-1} L^{-k} \int_{0}^{L} 
 x_{d}^{d-1} (L-x_{d})^{k-1-d} F(\frac{H}{2^{x_{d}}}) \frac{H}{2^{x_{d}}} \intd{x_{d}} 
\end{align}
By summing up for all days $d=1..k\!-\!1$, we get
 \begin{align*}
 AC_{k} &\geq k AC_{1} - 
2k(k-1) L^{-k} \sum_{d=1}^{k-1}
 \binom{k-2}{d-1}  \int_{0}^{L} 
 x_{d}^{d-1} (L-x_{d})^{k-1-d} F(\frac{H}{2^{x_{d}}}) \frac{H}{2^{x_{d}}} \intd{x_{d}} \\ 
 &\geq k AC_{1} - 
2k(k-1) L^{-k} 
\int_{0}^{L}  F(\frac{H}{2^{x}}) \frac{H}{2^{x}} 
\sum_{d=1}^{k-1}
 \binom{k-2}{d-1}
 x^{d-1} (L-x)^{(k-2)-(d-1)} \intd{x}
 \\ 
 &\geq k AC_{1} - 
2k(k-1) L^{-k} 
\int_{0}^{L}  F(\frac{H}{2^{x}}) \frac{H}{2^{x}} L^{k-2} \intd{x}
\\ 
&\geq k AC_{1} - 
2k(k-1) L^{-1} 2AC_{1} 
\geq k(1-\frac{4(k-1)}{\LValue}) \frac{H}{4(\LValue)} 
 \end{align*}
\end{proof}

\vspace{-0.5\baselineskip}
\section{Additional material from Section~\ref{sec:extys}}
\label{app:extys}
\vspace{-0.5\baselineskip}

\newcommand{\FBUI}[1]{F_{\varphi v}^{#1}} 

\newcommand{\nBUI}[1]{n_{\varphi v}^{#1}} 

\newcommand{\Fext}{F_{v_{i} \oplus \frac{\scfb{i}}{\scfa{i}}}}

\begin{Lemma}\cite{HartlineMS08OptimalMSoSN}
Consider a monotone submodular function $f:\buyers \to \reals$. Consider random set $I$ by 
choosing each buyer in $\buyers$ independently with probability at 
least $p$. Then $\expec[f (I)] \geq  p\cdot f (I)$.  
\label{stmt:SubmodEp}
\end{Lemma}

%
As utilities are quasilinear, the lowest quantity bought can only decrease if the price is scaled by $\alpha \geq 1$, but stays constant if the valuation is also scaled. 
%
\begin{Lemma}
For $\alpha \!\geq\! 1$, $\demandSet{}_{\alpha v}(p) \!=\! \demandSet{}_{v}(\frac{p}{\alpha})\!$ and 
$\!F_{v}(p) \leq  F_{\alpha v \oplus x}(p), \forall x \!\geq\! 0$. 
\label{stmt:FVIncr}
\end{Lemma}
%

\begin{proof}[of Lemma~\ref{stmt:FVIncr}]
Assume $F_{v}(p) > 0$ (otherwise second statement trivially holds) and let $S \in \demandSet{}_{v}(p)$ be a minimal preferred set (by $v$) at price $p$: $|S| = F_{v}(p)$.  

We claim that $F_{\alpha v \oplus x}(p) > 0$; otherwise $\emptyset \in \demandSet{}_{\alpha v \oplus x}(p)$ implying $0 \geq \alpha v(S) + x - p|S|  > v(S) - p|S|$,
contradicting $S \in \demandSet{}_{v}(p)$. 
Let $S_{\alpha} \in \demandSet{}_{\alpha v \oplus x}(p)$ be a minimal preferred set (by $\alpha v \oplus x$) at price $p$: $|S_{\alpha}| = F_{\alpha v \oplus x}$. 

Suppose towards a contradiction $F_{v}(p) > F_{\alpha v \oplus x}(p)$, implying $S_{\alpha} \not \in \demandSet{}_{v}(p)$. Thus, at price $p$, $S$ is strictly preferred by $v$ to $S_{\alpha}$ i.e. 
\begin{align} 
v(S) - p F_{v}(p) &> v(S_{\alpha}) - p F_{\alpha v \oplus x}(p) \text{ i.e. } \\
v(S)  - v(S_{\alpha}) &> p F_{v}(p) - p F_{\alpha v \oplus x}(p) > 0 \text{ implying } \\
\alpha v(S) - \alpha v(S_{\alpha}) &> p F_{v}(p) - p F_{\alpha v \oplus x}(p) \text{ i.e. } \\
\alpha v(S) + x - p F_{v}(p)  &> \alpha v(S_{\alpha}) + x - p F_{\alpha v \oplus x}(p) 
\end{align}
i.e. $S$ is strictly preferred to $S_{\alpha}$ by $\alpha v \oplus x$ at $p$, 
contradicting $S_{\alpha} \in \demandSet{}_{\alpha v \oplus x}(p)$.
%
\qedF \end{proof}

\noindent
Let $\inflADa{d}{i} = a_{i}(\inflProp_{d}\wo{i}), \inflADb{d}{i} = b_{i}(\inflProp_{d}\wo{i})$ be random variables for the influence   on buyer $i \in \buyers \!\setminus\!A_{1}$ just before day $d \in 1..k$ in $IE_{k}$. Thus $\inflADa{1}{i}=1, \inflADb{1}{i}=0$.

Lemma~\ref{stmt:FIncrHNoHGS} below
parallels Theorem~\ref{stmt:oidDS}. It lower bounds the 
quantity bought by a buyer outside $A_{1}$ from day $2$ up to a given day $d$. 


\begin{Lemma}
Fix buyer $i \in \buyers$ with $\inflADa{2}{i} \geq \scfa{i}, \inflADb{2}{i} \geq \scfb{i} $. 
Consider a price schedule $q_{x_{1}} > \ldots > q_{x_{k-1}}$ as in Def.~\ref{def:IEn} and fix day $d\in 2..k$. Assume that $\inflProp$ is monotone and each buyer's base valuation has hereditary maximizers. 

Let set $S_{d} \in \demandSet{S_{2}, \dots, S_{d-1}}_{\inflADa{d}{i} v_{i} \oplus \inflADb{d}{i}}(\scfa{i} \cdot q_{x_{1..d-1}})$ 
be preferred in day $d$ in $IE_{k}$ given (influenced) valuation $\inflADa{d}{i} v_{i} \oplus \inflADb{d}{i}$ and previously bought bundles $S_{2}, \!\dots,\! S_{d-1}\!$: 
Then $\sum_{\delta=2}^{d} |S_{\delta}| \geq \Fext(q_{x_{d-1}})$. 
\label{stmt:FIncrHNoHGS}
\end{Lemma}
\vspace{-0.75\baselineskip}

\newcommand{\utndGS}[2]{\inflADa{d}{i} v\left(#1 \cup \bigcup_{\delta=2}^{d-1} S_{\delta} \right) 
- \tst\scfa{i}  q_{x_{d-1}} (#2 + \nBUI{d-1}) - c_{d-1}
}

\begin{proof}[of Lemma~\ref{stmt:FIncrHNoHGS}]

\vspace{-2\baselineskip}
\begin{align*}
S_{d} \!\in\!\! \argmax_{S \cap \bigcup_{\delta=2}^{d-1} S_{\delta} = \emptyset}
\{ 
\inflADa{d}{i} v_{i}(S \cup \bigcup_{\delta=2}^{d-1} S_{\delta}) \!\oplus\! \inflADb{d}{i} 
- 
\scfa{i} \cdot q_{x_{d-1}} |S| - \sum_{\delta=2}^{d-1} \scfa{i} \cdot q_{x_{\delta-1}} |S_{\delta}| 
\} 
\end{align*}
\vspace{-0.5\baselineskip}


Assume wlog that the customer makes the first purchase in  day $2$ :  $S_{2} \neq \emptyset$.  


\noindent
By Lemma~\ref{stmt:FVIncr} ($\alpha = \frac{\inflADa{2}{i}}{\scfa{i} } \!\geq\! 1$ and $x \!=\! \frac{\inflADb{2}{i}}{\scfa{i}}$), 
$S_{2} \!\in\! 
\demandSet{}_{{\inflADa{2}{i}}v_{i} \oplus \inflADb{2}{i}} 
\left(\scfa{i}  \cdot q_{x_{1}} \right) =
\demandSet{}_{\!\frac{\inflADa{2}{i}}{\scfa{i}} v_{i} \oplus \frac{\inflADb{2}{i}}{\scfa{i}}} \!\!
\left( q_{x_{1}} \right) 
$ and 
$|S_{2}| \!\geq\! F_{\!v_{i} \oplus \frac{\inflADb{2}{i}}{\scfa{i}}\!}(q_{x_{1}}) \!\geq\! 
F_{\!v_{i} \oplus \frac{\scfb{i}}{\scfa{i}}\!}(q_{x_{1}})$. 
Lemma~\ref{stmt:FVIncr}  ($x \!=\!  \frac{\inflADb{2}{i}-\scfb{i}}{\scfa{i}}$) implies the second inequality  (holding with equality unless $F_{v_{i} \oplus \frac{\scfb{i}}{\scfa{i}}}(q_{x_{1}}) = 0$).

In day $d \!>\! 2$, $\bigcup_{\delta=2}^{d-1} S_{\delta} \!\neq\! \emptyset$; 
thus $\inflADb{d}{i}$ is added to $\inflADa{d}{i} v_{i}(S \cup \bigcup_{\delta=2}^{d-1} S_{\delta})$ in the argmax. 
We get 
$S_{d} \in 
\demandSet{S_{2}, \dots, S_{d-1}}_{v_{i} \oplus \frac{\inflADb{d}{i}}{\inflADa{d}{i}}}
\left(
\frac{\scfa{i}}{\inflADa{d}{i}}   q_{x_{1}}, \dots,  
\frac{\scfa{i}}{\inflADa{d}{i}}   q_{x_{d-2}}, 
\frac{\scfa{i}}{\inflADa{d}{i}}   q_{x_{d-1}}
 \right)$ 
 which equals
%
$\demandSet{S_{2}, \dots, S_{d-1}}_{v_{i} \oplus \frac{\scfb{i}}{\scfa{i}}}
\left(
\frac{\scfa{i}}{\inflADa{2}{i}}   q_{x_{1}}, \dots,  
\frac{\scfa{i}}{\inflADa{d-1}{i}} q_{x_{d-2}}, 
\frac{\scfa{i}}{\inflADa{d}{i}}   q_{x_{d-1}}
 \right)
$ since the current preferred set (see Eq.~\eqref{eq:seqUt}) 
is invariant to additions of scalars and to modifications of earlier prices 
(but not to current price, i.e. $\!\frac{\scfa{i}}{\inflADa{d}{i}} q_{x_{d-1}}\!$) 
given at least one earlier purchase. 
%
Clearly, since $v_{i}$  has hereditary maximizers, so does $v_{i}\oplus x$. 
Theorem~\ref{stmt:oidDS} for 
prices $\frac{\scfa{i}}{\inflADa{2}{i}} q_{x_{1}}, \dots,  
\frac{\scfa{i}}{\inflADa{d-1}{i}}   q_{x_{d-2}}, 
\frac{\scfa{i}}{\inflADa{d}{i}}  q_{x_{d-1}}$ 
yields 
$\sum_{\delta=2}^{d} |S_{\delta}| \!\geq\! \Fext(\frac{\scfa{i}}{\inflADa{d}{i}} q_{x_{d-1}})$. 
 As $\inflProp$, $a_{i}$ and $b_{i}$ are monotone, $\inflADa{d}{i} \!\geq\! \inflADa{2}{i} \!\geq\! \scfa{i}$.  
%
By Lemma~\ref{stmt:FVIncr} for $\alpha=\frac{\inflADa{d}{i}}{\scfa{i}}\geq 1$, $\Fext(\frac{\scfa{i}}{\inflADa{d}{i}} q_{x_{d-1}}) \geq \Fext(q_{x_{d-1}})$.
\end{proof}

By Lemma~2.1 in~\cite{HartlineMS08OptimalMSoSN}, re-stated above as 
Lemma~\ref{stmt:SubmodEp}, 
as $a_{i}, b_{i}$ are submodular and monotone and $\inflProp$ is monotone,  the \emph{expected} influence on a buyer $i$ after the day $1$ give-away in $IE_{k}$ is at least 
half the maximum influence. 
This implies that a constant fraction of such buyers are significantly influenced. 
\begin{Lemma}
For any buyer $i \!\in\! \buyers$, 
$\expec[\inflADa{2}{i}] \geq 0.5 H^{a} $ and 
$\expec[\inflADb{2}{i}] \geq 0.5 H^{b} $.
\label{stmt:inflpH}
Therefore 
$\prob[\inflADa{2}{i} \!\geq\! \scfa{i}] \geq \frac{1}{4}$  and $\prob[\inflADb{2}{i} \!\geq\! \scfb{i}] \geq \frac{1}{4}$. 
\label{stmt:inflProbLB}
\end{Lemma}
\begin{proof}
We only provide the proof for $\inflADa{2}{i}$ -- the one for $\inflADb{2}{i}$ is similar. 
Let  $ x = \prob[\inflADa{2}{i} \geq \scfa{i}]$.  
We have $\expec[\inflADa{2}{i}]  \leq x H^{a} + (1-x) \scfa{i} $. 
The claim follows via simple algebra from  
$\expec[\inflADa{2}{i}] \geq \frac{H^{a}}{2}$ (Lemma~\ref{stmt:inflpH}). 
\qedF\end{proof}
\vspace{-0.5\baselineskip}


 Finally, we prove Theorem~\ref{stmt:logn/kExtys}, i.e. $IE_{k}$'s $O(\frac{\log mn}{k})$-revenue approximation 
\begin{proof}[of Theorem~\ref{stmt:logn/kExtys}]
By Lemma~\ref{stmt:inflProbLB}, 
$\inflADa{2}{i} \geq \scfa{i}$ 
for a constant fraction of buyers outside  $A_{1}$.  
By Lemma~\ref{stmt:FIncrHNoHGS}, for any day $d = 2..k$, 
each such buyer $i \in \buyers \!\setminus\!A_{1}$
buys at least $\Fext(q_{x_{d-1}})$ items in total in days $2..d$
at prices $\scfa{i} \cdot q_{x_{1}},..,\scfa{i} \cdot q_{x_{d-1}}$.

For $q^{\delta} = \scfa{i} \cdot q_{x_{\delta-1}}$ 
and $x^{\delta} = \Fext(q_{x_{\delta-1}})$ for $\delta = 2..k$ 
in Lemma~\ref{stmt:groupSold} we get that buyer $i$ pays at least 
\begin{align}
\tst\sum_{\delta=2}^{k} \scfa{i} \cdot q_{x_{\delta-1}} (\Fext(q_{x_{\delta}}) - \Fext(q_{x_{\delta-1}})) 
\geq \scfa{i} \cdot \Omega(\frac{k}{\log \nOrmn}) (H_{i} + \frac{\scfb{i}}{\scfa{i}})   \notag
\end{align}
 by Theorem~\ref{stmt:logn/k}. 
The approximation factor follows after noting that the optimal marketing strategy can yield revenue at most $\sum_{i \in \buyers}  (H_{i} H^{a} + H^{b})$. 
\qedF\end{proof}

\end{document}

%% file: macrosFlo.tex
\usepackage{amssymb}

\newtheorem{Theorem}{Theorem}

\newtheorem{Definition}{Definition}
\newtheorem{Lemma}{Lemma}

\newtheorem{Example}{Example}

\newcommand{\intd}[1]{\mathrm{d}#1}
\newcommand{\all}{\ensuremath{\, \forall \,}}
\newcommand{\reals}{\ensuremath{\mathbb{R}}}
\newcommand{\integers}{\ensuremath{\mathbb{Z}}}

\newcommand*{\argmax}{\mathop{\mathrm{argmax}}} 

\newcommand{\tst}{\textstyle}

\newcommand{\expec}{\mathbb{E}}

\newcommand{\prob}{\mathbb{P}}

\newcommand{\sta}[1]{#1^{\ast}}

\newcommand{\wi}[1]{\ensuremath{\!\cup\!\{#1\}}}
\newcommand{\wo}[1]{\ensuremath{\!\setminus\!\{#1\}}}

\newcommand{\calF}{\ensuremath{\mathcal{F}}}
\newcommand{\calM}{\ensuremath{\mathcal{M}}}
\newcommand{\calP}{\ensuremath{\mathcal{P}}}

\newcommand{\HS}[1]{\ensuremath{\frac{H}{2^{#1}}}}
\newcommand{\scfa}[1]{{H^{a}\!}/{3}} 
\newcommand{\scfb}[1]{{H^{b}\!}/{3}} 
\newcommand{\HSEx}[1]{\ensuremath{\frac{H+H^{b}/H^{a}}{2^{#1}}}}

\newcommand{\os}[2]{x_{#1:#2}} 